\documentclass[12pt]{article}
\usepackage[margin=1in]{geometry}
\usepackage[small]{titlesec}
\usepackage{natbib,amsmath,amssymb,amsthm,graphicx,setspace,paralist,booktabs,rotating,subcaption,float,color}
\usepackage{times} 
\usepackage{multirow}
\usepackage[dvipsnames,table,dvipsnames*, svgnames*, hyperref]{xcolor}
\usepackage[colorlinks=true,linkcolor=blue,citecolor=blue,urlcolor=blue,breaklinks]{hyperref}
\usepackage{xr}
\usepackage{algpseudocode,float,algorithm}
   
\usepackage{amssymb,bm,bbm} 	
    
\AtBeginDocument{
\abovedisplayskip=5pt plus 2pt minus 2pt
\belowdisplayskip=\abovedisplayskip
\abovedisplayshortskip=2pt plus 2pt minus 2pt
\belowdisplayshortskip=\belowdisplayskip}
\titlespacing*{\section}{0pt}{*2}{*1}
\titlespacing*{\subsection}{0pt}{*2}{*1} 
\bibpunct{(}{)}{;}{a}{}{,}
\newtheorem{theorem}{Theorem}
\newtheorem{proc}{Procedure}
\newtheorem{corollary}{Corollary}
\newtheorem{proposition}{Proposition}

\newtheorem{lemma}{Lemma}
{
	\theoremstyle{definition}
	\newtheorem{definition}{Definition}
	\newtheorem{example}{Example}
	\newtheorem{remark}{Remark}
}
\newcommand{\beq}{\begin{equation}}
\newcommand{\eeq}{\end{equation}}
\newcommand{\beas}{\begin{align*}}
\newcommand{\eeas}{\end{align*}}
\newcommand{\bea}{\begin{align}}
\newcommand{\eea}{\end{align}}
\newcommand{\bei}{\begin{itemize}}
	\newcommand{\eei}{\end{itemize}}
\newcommand{\ben}{\begin{enumerate}}
	\newcommand{\een}{\end{enumerate}}
\newcommand{\bet}{\begin{theorem}}
	\newcommand{\eet}{\end{theorem}}
\newcommand{\bel}{\begin{lemma}}
	\newcommand{\eel}{\end{lemma}}
\newcommand{\bep}{\begin{proposition}}
	\newcommand{\eep}{\end{proposition}}
\newcommand{\bed}{\begin{definition}}
	\newcommand{\eed}{\end{definition}}
\newcommand{\bec}{\begin{corollary}}
	\newcommand{\eec}{\end{corollary}}
\newcommand{\bex}{\begin{example}}
	\newcommand{\eex}{\end{example}}

\newcommand{\Sig}{\bold{\Sigma}}

\newcommand{\bth}{\bm\theta}

\newcommand{\R}{\mathbb{R}}
\newcommand{\E}{\mathbb{E}}
\newcommand{\Pro}{\mathbb{P}}

\newcommand{\HH}{\mathcal{H}}

\newcommand{\argmin}{\mathop{\rm arg\min}}
\newcommand{\argmax}{\mathop{\rm arg\max}}

\numberwithin{equation}{section}

\def\Var{{\rm Var}}

\def\xx{\bm{x}}

\newcommand{\bbeta}{\bm{\beta}}

\allowdisplaybreaks[1]

\newcommand{\ignore}[1]{}
\begin{document}

\begin{titlepage}
\setstretch{1.24}
\title{Estimation, Confidence Intervals, and Large-Scale Hypotheses Testing for High-Dimensional Mixed Linear Regression}
	\author{Linjun Zhang, \ Rong Ma, \ T. Tony Cai, \ and \ Hongzhe Li}
	\date{}
	\maketitle
	\thispagestyle{empty}
	\footnotetext{Linjun Zhang is Assistant Professor, Department of Statistics, Rutgers University, Piscataway, NJ 08854. (E-mail: \emph{linjun.zhang@rutgers.edu}). Rong Ma is PhD Candidate, Department of Biostatistics, Epidemiology and Informatics, Perelman School of Medicine, University of Pennsylvania, Philadelphia, PA 19104 (E-mail: \emph{rongm@pennmedicine.upenn.edu}). T. Tony Cai is Daniel H. Silberberg Professor of Statistics, Department of Statistics, The Wharton School, University of Pennsylvania, Philadelphia, PA 19104 (E-mail:\emph{tcai@wharton.upenn.edu}). Hongzhe Li is Professor of Biostatistics and Statistics, Department of Biostatistics, Epidemiology and Informatics, Perelman School of Medicine, University of Pennsylvania, Philadelphia, PA 19104 (E-mail: \emph{hongzhe@upenn.edu}).}

\begin{abstract}
	This paper studies the high-dimensional  mixed linear regression (MLR) where the output variable comes from one of the two linear regression models with an unknown mixing proportion and an unknown covariance structure of the random covariates. Building upon a high-dimensional EM algorithm, we propose an iterative procedure for estimating the two regression vectors and establish their rates of convergence. 
	Based on the iterative estimators, we further construct debiased estimators and establish their asymptotic normality. For individual coordinates, confidence intervals centered at the debiased estimators are constructed. 

	Furthermore, a large-scale multiple testing procedure is proposed for testing the regression coefficients and is shown to control the false discovery rate (FDR) asymptotically. Simulation studies are carried out to examine the numerical performance of the proposed methods and their superiority over existing methods. The proposed methods are further illustrated through an analysis of a dataset of multiplex image cytometry, which investigates the interaction networks among the cellular phenotypes that include the expression levels of 20  epitopes or combinations of markers. 

\bigskip
\noindent\emph{KEYWORDS}: debiasing, EM algorithm, FDR, iterative estimation, large-scale multiple testing
\end{abstract}

\end{titlepage}

\section{INTRODUCTION}

Mixed linear regression (MLR) models are widely used in analyzing heterogeneous data arising from biology, physics, economics, and business  \citep{mclachlan2004finite, grun2007applications, netrapalli2013phase, li2019drug, devijver2020clustering}. In many of these modern applications, the number of the covariates is comparable with, or sometimes far exceeds, the number of observed samples. In such high-dimensional settings, statistical inference methods designed for estimation and hypothesis testing of the regression coefficients in the classical low-dimensional setting are often not valid. 
There is a paucity of methods and fundamental theoretical understanding on  statistical estimation and inference for high-dimensional  MLR models.  This motivates us to develop computationally efficient and theoretically guaranteed statistical methods for high-dimensional MLR models in analyzing  large heterogeneous datasets. 

We consider the following high-dimensional MLR model where the observed data are $i.i.d.$ draws from one of the two unknown linear models: 
\beq\label{eq:model} 
y_i=\left\{ \begin{array}{ll}
\xx_i^\top\bbeta^*_1+\epsilon_i  &\textrm{with probability $\omega^*$,}\\
\xx_i^\top\bbeta^*_2+\epsilon_i  &\textrm{with probability $1-\omega^*$,}
\end{array} \right.
\quad i=1,2,\ldots, n,\quad \bbeta_1^*, \bbeta_2^*\in \R^p,
\eeq
where  the random design variables $\xx_i\in\R^p$ are $i.i.d.$ samples from $N_p(0,\Sig)$ with the unknown covariance matrix $\Sig$, 
 $\bbeta^*_1\ne \bbeta^*_2$ are latent regression coefficients, $\omega^*\in(0,1)$ is the  unknown mixing proportion, and the noise $\epsilon_i$ is $i.i.d.$ from $N(0,\sigma^2)$ for some $\sigma>0$. For identifiability, we assume $\omega^*\in(1/2,1)$. Under the high-dimensional setting where $p$ is much larger than $n$, we aim to  answer the following inference questions:
\begin{enumerate}
	\item What is an efficient algorithm for estimating the underlying regression vectors $\bbeta_1^*$ and $\bbeta_2^*$ where both the mixing proportion $\omega^*$ and the random design covariance matrix $\Sig$ are unknown?  What is the rate of convergence of the estimator?
	\item How to construct asymptotically valid tests and {confidence intervals} for the individual coordinates of the latent regression coefficients $\bbeta^*_1$ and $\bbeta_2^*$, and their difference $\bbeta^*_1-\bbeta_2^*$?
	\item For simultaneously testing the null hypotheses $H_{0j}: \bbeta_{1j}^*=\bbeta_{2j}^*=0$, $j=1,...,p$, how to construct a large-scale multiple testing procedure that controls the false discovery rate (FDR)  and false discovery proportion (FDP) asymptotically?
\end{enumerate}

\subsection{Related Works}

In the classical low-dimensional settings, the problems of estimation, hypotheses testing and confidence intervals for MLR have been extensively studied in literature. For example, \cite{zhu2004hypothesis} considered hypothesis testing and developed an asymptotic theory for both the maximum likelihood and the maximum modified likelihood estimators in MLR. \cite{khalili2007variable} introduced a penalized likelihood approach for variable selection in MLR. \cite{faria2010fitting} compared three expectation-maximization (EM) algorithms that compute the maximum likelihood estimates of the coefficients of MLR. \cite{chaganty2013spectral} developed a  computationally efficient algorithm based on the tensor power method, and obtained the rates of convergence for their proposed estimator. \cite{bashir2012robust} proposed a robust model that can achieve high breakdown point in the contaminated data for parameter estimation in MLR. Based on a new initialization step for the EM algorithm, \cite{yi2013alternating} provided the theoretical guarantees for coefficient estimation in MLR. Moreover, \cite{yao2015mixtures} proposed a deconvolution method to study the MLR with measurement errors. \cite{zhong2016mixed} proposed a non-convex continuous objective function for solving the general unbalanced $k$-component MLR.

\cite{balakrishnan2017statistical} developed a general framework for proving rigorous guarantees on the performance of the EM algorithm and applied to some statistical problems including the estimation of coefficients in the symmetric MLR where the mixing proportion  is known to be $\omega^*=1/2$. \cite{li2018learning} presented a fixed parameter algorithm that solves MLR under Gaussian design in time that is nearly linear in the sample size and the dimension. More recently, building upon the work of \cite{balakrishnan2017statistical} on the symmetric MLR, \cite{mclachlan2004finite} and \cite{klusowski2019estimating} introduced better tools for analyzing the convergence rates of the EM algorithm for estimating the coefficients. \cite{shen2019iterative} proposed an efficient algorithm, Iterative Least Trimmed Squares, for solving MLR with adversarial corruptions.

In contrast, statistical inference for MLR in the high-dimensional setting is relatively less studied. Specifically, \citet{stadler2010l1} proposed an $\ell_1$ penalized estimator and developed an efficient EM algorithm for MLR with provable convergence properties. \cite{wang2014high} and \cite{yi2015regularized} established a general theory of the EM algorithm for statistical inference in high dimensional latent variable models, including the high-dimensional MLR with the symmetric and spherical assumptions. In \cite{zhu2017high}, a generic stochastic EM algorithm was proposed for the high-dimensional MLR with theoretical guarantees obtained under the symmetric setting ($\omega^*=1/2$). More recently, \cite{fan2018curse} studied the fundamental tradeoffs between statistical accuracy and computational tractability for high-dimensional latent variables models, including testing the global null hypothesis in MLR. However, problems such as statistical inference about the individual regression coefficients and large-scale multiple testing under the general MLR with an unknown mixing proportion and an unknown design covariance matrix have not been addressed in the literature.

\subsection{Main Contributions}

The main contributions of our paper are three-fold.
\begin{enumerate}
	\item Based on a careful analysis of a high-dimensional EM algorithm, we propose iterative estimators for the regression coefficients $(\bbeta_1^*,\bbeta_2^*)$ without the knowledge of the mixing proportions or the design covariance matrix, and obtain explicitly the rates of convergence of the iterative estimators under the $\ell_2$ norm. To the best of our knowledge, this is the first result on the estimation of the high-dimensional MLR with both unknown mixing proportion and unknown design covariance matrix.
	\item
Further, we construct debiased estimators of the latent regression coefficients, based on the iterative estimators, and establish the asymptotic normality of its individual coordinates. The limiting distribution is then used for constructing confidence intervals and tests for the individual latent regression coefficients.
	\item For the problem of large-scale testing of hypotheses $H_{0j}: \bbeta_{1j}^*=\bbeta_{2j}^*=0$, $j=1,...,p$, we propose a multiple testing procedure that is  shown to control the FDR and FDP asymptotically. Strong numerical results suggest the superior empirical performance of our proposed testing procedure  over the existing methods.
\end{enumerate}

\subsection{Organization and Notation}
Throughout our paper, for a vector $\bold{a} = (a_1,...,a_n)^\top \in \mathbb{R}^{n}$, we define the $\ell_p$ norm $\| \bold{a} \|_p = \big(\sum_{i=1}^n a_i^p\big)^{1/p}$, and the $\ell_\infty$ norm $\| \bold{a}\|_{\infty} = \max_{1\le j\le n}  |a_{i}|$. $\bold{a}_{-j}\in \R^{n-1}$ stands for the subvector of $\bold{a}$ without the $j$-th component. For vectors $\bold{a},\bold{b}\in \R^n$, we denote their inner product $\langle \bold{a},\bold{b} \rangle=\sum_{i=1}^na_ib_i$. For a matrix $A\in \R^{p\times q}$, $\lambda_i(A)$ stands for the $i$-th largest singular value of $A$ and $\lambda_{\max}(A)  = \lambda_1(A)$, $\lambda_{\min}(A) = \lambda_{p \wedge q} (A)$. $\|A\|_1$ denotes the matrix $\ell_1$ norm, and $\|A\|_\infty=\max_{i,j}|A_{ij}|$. In addition, $A_{-i.-j}\in \R^{(p-1)\times (q-1)}$ stands for the submatrix of $A$ without the $i$ th row and $j$-th column. For any positive integer $p$, we denote $[p]=\{1,...,p\}$. Furthermore, for sequences $\{a_n\}$ and $\{b_n\}$, we write $a_n = o(b_n)$ if $\lim_{n} a_n/b_n =0$, and write $a_n = O(b_n)$, $a_n\lesssim b_n$ or $b_n \gtrsim a_n$ if there exists a constant $C$ such that $a_n \le Cb_n$ for all $n$. We also write $a_n = O_P(b_n)$ if there exists a constant $C$ such that $\lim\inf_{n\to\infty}\Pro(a_n\le C b_n)=1$, and  $a_n = o_P(b_n)$ if $a_n/b_n\stackrel{p}{\to} 1$.
 We write $a_n\asymp b_n$ if $a_n \lesssim b_n$ and $a_n\gtrsim b_n$. For a set $A$, we denote $|A|$ as its cardinality. Lastly, $C, C_0, C_1,...$ are constants that may vary from place to place.

The rest of the paper is organized as follows.  We propose in Section 2 the iterative algorithm and the estimators of the latent regression coefficients and study their theoretical properties. Section 3 introduces the debiased estimators of individual regression coefficients and obtains their asymptotic normality and the resulting confidence intervals. In Section 4, by focusing on the problem of testing  large-scale simultaneous hypotheses, we present our multiple testing procedure and show that it controls the FDR/FDP asymptotically. In Section 5, the numerical performance of the proposed methods are evaluated through extensive simulations. In Section 6, the proposed procedures are illustrated by an analysis of a multiplex image cytometry dataset. Further extensions and related problems are discussed in Section 7. The proofs of other theorems as well as technical lemmas are collected in the Supplementary Materials  \citep{supp}.

\section{ITERATIVE ESTIMATION VIA THE EM ALGORITHM}

Suppose we have $n$ observations $\{(\bm x_i ,y_i)\}_{i=1}^n$ generated independently from the MLR model in \eqref{eq:model}, and wish to estimate and make inference on the coefficient vectors $\bbeta_1^*$ and $\bbeta_2^*$. In the classical setting where $p$ is fixed or much smaller than $n$, the maximum likelihood estimator (MLE) has been shown to perform well under mild conditions \citep{balakrishnan2017statistical}. The MLE aims to maximize the log-likelihood of the data $\{(\bm x_i ,y_i)\}_{i=1}^n$, which can be written as

\begin{equation}\label{eq:likelihood}
l_n(\bth;\xx,y)=\frac{1}{n}\sum_{i=1}^n\log\bigg[ \frac{\omega}{\sqrt{2\pi}\sigma}\exp\left\{ {-\frac{(y_i-\langle \xx_i,\bbeta_1\rangle )^2}{2\sigma^2}}\right\}+  \frac{1-\omega}{\sqrt{2\pi}\sigma}\exp\left\{ {-\frac{(y_i-\langle \xx_i,\bbeta_2\rangle )^2}{2\sigma^2}}\right\}\bigg],
\end{equation}
where we denote the parameter $\bth=(\omega,\bbeta_1,\bbeta_2)$ and the log-likelihood by $l_n(\bth)$.

Due to the non-convexity of $l_n(\bth;\xx,y)$,  searching for the MLE is computationally intractable. 
Moreover, in the high-dimensional setting where the dimension $p$ is much larger than the sample size $n$, the MLE is in general not well defined, unless the models are carefully regularized by sparsity-type
assumptions. In this paper, we propose to explore the sparsity of the coefficient vectors. Further, we develop an  EM algorithm to address the extra computational challenge for parameter estimation and uncertainty assessment. 

\subsection{High-Dimensional EM Algorithm and the Iterative Estimators}

For ease of presentation, let us use $z_i$ to denote  the hidden labels of $(\bm x_i ,y_i)$, that is, $z_i=1$ if $(\bm x_i ,y_i)$ is drawn from the first model $y_i=\bm x_i^\top\bbeta_1^*+\epsilon_i$, and $z_i=2$ if the  underlying truth is the second model $y_i=\bm x_i^\top\bbeta_2^*+\epsilon_i$. The marginal distribution of $z_i$ is given by $\Pro(z_i=1)=1-\Pro(z_i=2)=\omega$.
The EM algorithm is essentially an alternating maximization method, which alternatively optimizes between the identification of hidden labels $\{z_i\}_{i=1}^n$ and the estimation of parameter $\bth=(\omega,\bbeta_1,\bbeta_2)$. 

Specifically, in the E-step of $(t+1)$-th iteration, given the parameters $\bth^{(t)}=(\omega^{(t)},\bbeta_1^{(t)},\bbeta_2^{(t)})$ estimated from the previous $t$-th step, the conditional probability of the $i$-th sample in class 1 given the observed data $(\bm x_i ,y_i)$ can be calculated as
\begin{equation}\label{eq:gamma}
\gamma_{\bth^{(t)}}(\xx_i,y_i):=\Pro_{\bth^{(t)}}(z_i=1\mid\bm x_i,y_i)=\frac{\omega^{(t)}\exp(-\frac{(y_i-\langle\xx_i,\bbeta_1^{(t)}\rangle)^2}{2\sigma^2})}{\omega^{(t)}\exp(-\frac{(y_i-\langle\xx_i,\bbeta^{(t)}_1\rangle)^2}{2\sigma^2})+(1-\omega^{(t)})\exp(-\frac{(y_i-\langle\xx_i,\bbeta_2^{(t)}\rangle)^2}{2\sigma^2})}.
\end{equation}
As a  result, the conditional expectation of the log-likelihood \eqref{eq:likelihood}, with respect to the conditional distribution given $(\bm x, y)$ under the current estimate of the parameter $\bth^{(t)}$, can be calculated as
\begin{align}\label{eq:Q}
Q_n(\bth\mid\bth^{(t)}):=&
\E_{\bth^{(t)}}[l_n(\bth;\xx,y)\mid \xx, y]\\
\notag=&-\frac{1}{2n}\left[\sum_{i=1}^n\gamma_{\bth^{(t)}}(\xx_i,y_i)(y_i-\langle \xx_i,\bbeta_1\rangle )^2+\sum_{i=1}^n(1-\gamma_{\bth^{(t)}}(\xx_i,y_i))(y_i-\langle \xx_i,\bbeta_2\rangle )^2\right]\\
&\notag+\frac{1}{n}\sum_{i=1}^n(1-\gamma_{\bth^{(t)}}(\xx_i,y_i))\log(1-\omega)+\gamma_{\bth^{(t)}}(\xx_i,y_i)\log\omega .
\end{align}
Given $\gamma_{\bth^{(t)}}(\xx_i,y_i)$, i.e., the distribution of the latent labels, the M-step is usually proceeded by maximizing  $Q_n(\bth\mid\bth^{(t)})$: $$
\hat\bth^{(t+1)}=\argmax_{\bth}  Q_n(\bth\mid\bth^{(t)}).
$$
However, in the high-dimensional setting, such a maximization tends to overfit data. To handle the
challenge of high-dimensionality, the key ingredient of our algorithm is to add a regularization term $\|\bbeta\|_1$ to enforce sparsity. In particular, we write $\gamma^{(t)}_{\bm \theta,i}=\gamma_{\bth^{(t)}}(\xx_i,y_i)$, and let
\begin{equation}\label{eq:beta1}
\begin{aligned}
\hat\bbeta_1^{(t+1)}&=\argmin_{\bbeta_1}\frac{1}{2n}\sum_{i=1}^n\gamma^{(t)}_{\bm \theta,i}(y_i-\langle \xx_i,\bbeta_1\rangle )^2+\lambda_n^{(t+1)} \|\bbeta_1\|_1\\
\hat\bbeta_2^{(t+1)}&=\argmin_{\bbeta_2}\frac{1}{2n}\sum_{i=1}^n(1-\gamma^{(t)}_{\bm \theta,i})(y_i-\langle \xx_i,\bbeta_2\rangle )^2+\lambda_n^{(t+1)} \|\bbeta_2\|_1,
\end{aligned}
\end{equation}
where $\lambda^{(t+1)}$ is a tuning parameter which will also be updated recursively, and will be specified later. 
We also update $\omega^{(t+1)}$ by $$
\omega^{(t+1)}=\frac{1}{n}\sum_{i=1}^n \gamma_{\bth^{(t)}}(\xx_i,y_i).
$$

Given a suitable initialization, the proposed high-dimensional EM algorithm then proceeds by iterating between the E-step and the M-step, which is summarized in the following Algorithm \ref{alg:hd}.

\begin{algorithm}
	\caption{EM for High-Dimensional MLR}\label{alg:hd}
	\begin{algorithmic}[1]
		\State {\bf Inputs:} Initializations $\hat\omega^{(0)}, \hat\bbeta_1^{(0)}, \hat\bbeta_2^{(0)}$, maximum number of iterations $T$, and constants $\kappa\in(0,1)$, $C_\lambda>0$. Split the Dataset into $T$ subsets of size $n/T$. For $i\in[n]$, set 
		$$
		\gamma^{(0)}_{\bm \theta,i}=\frac{\hat\omega^{(0)}\exp(-\frac{(y_i-\langle\xx_i,\hat\bbeta^{(0)}_1\rangle)^2}{2\sigma^2})}{\hat\omega^{(0)}\exp(-\frac{(y_i-\langle\xx_i,\hat\bbeta^{(0)}_1\rangle)^2}{2\sigma^2})+(1-\hat\omega^{(0)})\exp(-\frac{(y_i-\langle\xx_i,\hat\bbeta^{(0)}_2\rangle)^2}{2\sigma^2})}.
		$$
		\For {$t=0,1,\ldots,T-1$}
		\State {\bf E-Step:} Evaluate $Q_n(\bth \mid \hat\bth^{(t)})$ as defined in \eqref{eq:Q} with the $t$-th data subset.
		\State {\bf M-Step:} Update $\bbeta_1^{(t+1)}$ and $\hat\bbeta^{(t+1)}$ via 
		\begin{equation}\label{eq:beta2}
		\begin{aligned}
		\hat\bbeta_1^{(t+1)}&=\argmin_{\bbeta_1}\frac{1}{2n}\sum_{i=1}^n\gamma^{(t)}_{\bm \theta,i}(y_i-\langle \xx_i,\bbeta_1\rangle )^2+\lambda_n^{(t+1)} \|\bbeta_1\|_1\\
		\hat\bbeta_2^{(t+1)}&=\argmin_{\bbeta_2}\frac{1}{2n}\sum_{i=1}^n(1-\gamma^{(t)}_{\bm \theta,i})(y_i-\langle \xx_i,\bbeta_2\rangle )^2+\lambda_n^{(t+1)} \|\bbeta_2\|_1,
		\end{aligned}
		\end{equation}
		with 
		\beq\label{lambda:alg1}
		\lambda_n^{(t+1)}=\kappa_{\lambda} \lambda_n^{(t)}+C_{\lambda}\sqrt\frac{\log p}{n}.
		\eeq
		Update $\hat\omega^{(t)}$ via $
		\hat\omega^{(t)}=\frac{1}{n}\sum_{i=1}^n\gamma^{(t)}_{\bm \theta,i}.$
		\EndFor
		\State Output $\bbeta_1^{(T)}$ and $\hat\bbeta_2^{(T)}$.
	\end{algorithmic}
\end{algorithm}

\begin{remark}
	In the above algorithm, it is required that the noise level $\sigma^2$ is known. Such an assumption is widely used in prior literature in mixed linear regressions, see \cite{wang2014high, balakrishnan2017statistical, klusowski2019estimating} and reference therein. In practice, a good estimator of $\sigma^2$ can be substituted in the algorithm to achieve deisrable empirical performance. See Section \ref{sec:sim} for more detailed numerical justifications.

%
\end{remark}

\subsection{Rate of Convergence}

In this section, we give theoretical guarantees for estimating the coefficient vectors $\bbeta_1^*$ and $\bbeta_2^*$ using Algorithm \ref{alg:hd}. To begin with, we introduce the parameter space for $(\omega^*,\bbeta_1^*,\bbeta_2^*)$, where we assume that $\bbeta_1^*$ and $\bbeta_2^*$ are both sparse vectors and $\omega^*$ is bounded away from 0 or 1. 
\[
\Theta(s) = \bigg\{(\omega,\bbeta_1,\bbeta_2):  \omega\in (c,1-c), \|\bbeta_1\|_0,\|\bbeta_2\|_0\le s, \text{ for some } c\in(0,1/2) \bigg\}.
\]
{Furthermore, we introduce the following regularity conditions on the initialization and signal-to-noise ratio (SNR) strength. 
	\begin{itemize}
		\item[(A1)]: Initialization: $\|\bbeta_1^{(0)}-\bbeta_1^*\|_2+\|\bbeta_2^{(0)}-\bbeta_2^*\|+|\omega^{(0)}-\omega^*|\le\min\{\omega^*/2, (1-\omega^*)/2, c_l\cdot{\Delta^*}\}$, where $\Delta^*=\sqrt{(\bbeta_1^*-\bbeta_2^*)^\top\Sig^{-1}(\bbeta_1^*-\bbeta_2^*)}$;
		\item[(A2)]: SNR strength: $(\bbeta_1^*-\bbeta_2^*)^\top\Sig^{-1}(\bbeta_1^*-\bbeta_2^*)\ge c_s$,
	\end{itemize}
	where $c_l, c_s>0$ are some universal constants that and do not grow with $n$ or $p$.
}
The (A1) suggests the initialized estimator should be closed to the truth. Such a condition is common in the literature of mixed linear regression, see \cite{balakrishnan2017statistical, yi2013alternating, wang2014high, yi2015regularized}. In practice, our initialization algorithm is discussed in Section~\ref{sec:sim}. Condition (A2) has also been commonly used in the literature of mixed linear regression \citep{balakrishnan2017statistical,klusowski2019estimating}, and other hidden variable models \citep{wang2014high,cai2019chime}.

\begin{theorem}\label{thm:estimation}
	Suppose $1/M<\lambda_{\min}(\Sig)\le \lambda_{\max}(\Sig)<M$ for some constant $M>1$ and conditions (A1) and (A2) hold. 
	{If {$\frac{s\log p{\cdot\log n}}{n} = o(1)$}, and $c_l$ is sufficiently large, that is, $c_l\ge C(\omega^*, M, c_l)$ where $C(\omega^*, M, c_l)$ is a constant depending only on $(\omega^*, M, c_l)$. Then $T\gtrsim \log n$}, we have with probability at least $1-p^{-1}$,
	$$
	\|\hat\bbeta_1^{(T)}-\bbeta_1^*\|_2+\|\hat\bbeta_2^{(T)}-\bbeta_2^*\|_2\lesssim\sqrt\frac{s\log p{\cdot\log n}}{n}.
	$$
\end{theorem}

\begin{remark}
The condition on the spectrum of $\Sig$ is standard in the high-dimensional literature. For example, it has been used in \citet{cai2016estimating, cai2012optimal} and \citet{javanmard2014confidence} for estimation of precision matrices, covariance matrices and regression coefficients, respectively. Further, the convergence rate of optimization error is exponentially fast, so we only need that the number of iterations $T\gtrsim \log n$ to make the optimization error negligible comparing to the statistical error.
\end{remark}


\section{DEBIASED ESTIMATORS AND THEIR ASYMPTOTIC NORMALITY}

The iterative estimators obtained from the high-dimensional EM algorithm (Algorithm \ref{alg:hd}) enjoys desirable properties in term of squared error, they are however unsuitable to be used directly for statistical inference. In this section, we introduce the debiased estimators for the mixed linear regression coefficients $\bbeta^*_{\ell j}$ with $\ell\in\{1,2\}$ and $j\in[p]$, and obtain their asymptotic normality, which can then be used to perform hypothesis testing and construct confidence intervals for the individual coefficients. 

\subsection{Debiased Estimators}

Due to the $\ell_1$ regularization in the M-step, the outputs $\hat\bbeta_1^{(T)}$ and $\hat\bbeta_2^{(T)}$ from the high-dimensional EM algorithm (Algorithm \ref{alg:hd})  are biased. To facilitate the subsequent statistical inference, we proceed by correcting their biases. Such a de-biased procedure has been used widely in high-dimensional single linear regression models \citep{javanmard2014confidence,javanmard2014hypothesis,van2014asymptotically,zhang2014confidence,ning2017general}, but cannot be directly applied to the EM solutions. In the following, we first present some high-level intuition. 

We start with the regression coefficient $\bbeta^*_1$. Note that in Algorithm~\ref{alg:hd}, $\hat\bbeta_1^{(T)}$ is constructed only based on the $T$-th sample, and the sample size is $n/T$ with $T\asymp \log n$. In the following, for the notational simplicity, we simply write $n_T=n/T$.
Firstly, $\hat\bbeta_1^{(T)}$ satisfies the Karush-Kuhn-Tucker (KKT) condition
\begin{equation}\label{kkt}
-\frac{1}{n}\sum_{i=1}^{n_T}\gamma_{\bm\theta,i}^{(T)}(y_i-\langle\xx_i,\hat\bbeta_1^{(T)}\rangle)\xx_i+\lambda_n^{(T)} \partial ||\hat\bbeta_1^{(T)}||_1=0,
\end{equation}
where  $\partial ||\hat\bbeta_1^{(T)}||_1$ is the subgradient of the $\ell_1$ norm $ ||\cdot||_1$.
Letting $\widehat\Sigma_{XX}=\frac{1}{n_T}\sum_{i=1}^{n_T}\gamma_{\bm\theta,i}^{(T)}\xx_i\xx_i^\top$ and $\widehat\Sigma_{XY}=\frac{1}{n_T}\sum_{i=1}^{n_T}\gamma_{\bm\theta,i}^{(T)}\xx_iy_i$, equation \eqref{kkt} can then be rewritten as
$$
\widehat\Sigma_{XX}\hat\bbeta_1^{(T)}-\widehat\Sigma_{XY}+\lambda_n^{(T)} \partial ||\hat\bbeta_1^{(T)}||_1=0,
$$
and as a result, 
$$
\widehat\Sigma_{XX}(\hat\bbeta_1^{(T)}-\bm\beta_1^*)+\lambda_n^{(T)} \partial ||\hat\bbeta_1^{(T)}||_1=\widehat\Sigma_{XY}-\widehat\Sigma_{XX}\bm\beta_1^*.
$$
Following the debiased Lasso method in  \citet{javanmard2014confidence}, suppose one has a good approximation of the ``inverse" of $\widehat\Sigma_{XX}$, say $M$, then one can multiply $M$ on the left to obtain
 $$
M\widehat\Sigma_{XX}(\hat\bbeta_1^{(T)}-\bm\beta_1^*)+\lambda_n^{(T)} M\partial ||\hat\bbeta_1^{(T)}||_1=M(\widehat\Sigma_{XY}-\widehat\Sigma_{XX}\bm\beta_1^*).
$$
Then it follows\begin{equation}\label{decomposition}
(\hat\bbeta_1^{(T)}+\lambda_n^{(T)} M\partial ||\hat\bbeta_1^{(T)}||_1)-\bm\beta_1^*=M(\widehat\Sigma_{XY}-\widehat\Sigma_{XX}\bm\beta_1^*)+(I-M\widehat\Sigma_{XX})(\hat\bbeta_1^{(T)}-\bm\beta_1^*).
\end{equation}
By inspection, if we let $\widehat{\bm \beta}_1^u=\hat\bbeta_1^{(T)}+\lambda_n^{(T)} M\partial ||\hat\bbeta_1^{(T)}||_1$, then 
\begin{align}\label{decomposition2}
\notag\sqrt n(\widehat{\bm \beta}_1^u- \bbeta_1^{*})= &\sqrt n(M\widehat\Sigma_{XY}-M\widehat\Sigma_{XX}{\bm \beta_1^*})+\sqrt n(I-M\widehat\Sigma_{XX})(\hat{\bm \beta}_1^{(T)}-{\bm \beta}_1^*)\\
\notag=&\sqrt n\bigg[\frac{1}{n}\sum_{i=1}^n\gamma_{\bm\theta,i}^{(T)}(y_i-\langle\xx_i,\bbeta_1^*\rangle)M\xx_i\bigg]+ o_P(1),
\end{align}
where the second equality incorporated the assumption that $M$ approximate the ``inverse" of $\widehat\Sigma_{XX}$ well and thus {$\|(I-M\widehat\Sigma_{XX})(\hat\bbeta_1^{(T)}-\bm\beta_1^*)\|_\infty\le\|I-M\widehat\Sigma_{XX}\|_\infty\|\hat\bbeta_1^{(T)}-\bm\beta_1^*\|_1$ is negligible}.

Unlike the procedure in  \citet{javanmard2014confidence}, our $\hat\Sigma_{XX}$ depends on $(\xx_i,y_i)$ instead of only on $\xx_i$'s, and therefore solving a direct approximation will mess up with the subsequent asymptotic normality. We propose to solve $M$ by the following two-step procedure. First,  let $\tilde\Sigma_{XX}=\frac{1}{n}\sum_{i=1}^n\xx_i\xx_i^\top$, for $j\in[p]$, let $\bm{\tilde m}_j$ be the solution of 
\begin{equation}\label{opt}
\begin{aligned}
& \underset{\bm m_j\in \R^p}{\text{minimize}}
& &\bm m_j^\top\hat\Sigma_{XX} \bm m_j\\
& \text{subject to}
& & ||\hat\Sigma_{XX} \bm m_j- e^{(p)}_j||_\infty\le \mu,\\
& & &\|\bm m_j\|\le C\sqrt{\log n}.
\end{aligned}
\end{equation}
where $\mu$ and $C$ are tuning parameters that will be discussed later. 

Second we set $
\bm m_j=\bm{\tilde m}_j/\hat\omega^{(T)},$ with $\hat\omega^{(T)}=\hat\omega^{(T)}\frac{1}{n}\sum_{i=1}^n\gamma_{\bm\theta,i}^{(T)}$. The denominator is used because $\tilde\Sigma_{XX}$ is approximately $\omega^*\cdot\hat\Sigma_{XX}$. Although $\hat\omega^{(T)}$ still depends on $(\xx_i,y_i)$, but it is close to $\omega^*$ and the distance is negligible. 

Now, to find the asymptotic distribution of $\sqrt{n_T}(\widehat{\bm \beta}_1^u- \bbeta_1^{*})$, let us consider the dominating term  \beq\label{eq:distribution}
\frac{1}{\sqrt{n_T}}\sum_{i=1}^{n_T}\gamma_{\bm\theta,i}^{(T)}(y_i-\langle\xx_i,\bbeta_1^*\rangle)M\xx_i.
\eeq
By inspection, conditioning on $\xx$, \eqref{eq:distribution} is an approximation of the linear transformation of the score function $\nabla_{\bth^*} l_n(\bth^*;\xx,y)$. Specifically, straightforward computation yields the score function
\beq\label{eq:score}
\nabla_{\bbeta_1} l_n(\bth^*;\xx,y)=\frac{1}{n_T}\sum_{i=1}^{n_T}\gamma_{\bth^*,i}(y_i-\langle \xx_i,\bbeta_1^* \rangle)\xx_i.
\eeq
By Theorem~\ref{thm:estimation}, $\bth^{(T)}$ is close to $\bth^*$, so heuristically, \eqref{eq:distribution} is also close to a linear transformation of the score function \eqref{eq:score} when $n$ is large. Moreover, since the score function  $\nabla l_n(\bth;\xx,y)$ is asymptotically normal  at the truth $\bth=\bth^*$ with covariance matrix being the information matrix $I(\bth^*)=-\E_{\bth^*}\nabla^2 l_n(\bth^*;\xx,y)$, in order to make valid inference about the parameters, it suffices to estimate the information matrix. 

The following Lemma~\ref{lemma:score} provides the an estimator for information matrix and the corresponding asymptotic distribution of \eqref{eq:distribution}. 

\begin{lemma}\label{lemma:score}
	Recall $\bth=(\omega,\bbeta_1,\bbeta_2)$, and denote $	T_n(\bth)=-\left(\nabla^2_{\bth} Q_n(\bth\mid\bth')+\nabla^2_{\bth,\bth'} Q_n(\bth\mid\bth')\mid_{\bth,\bth'=\bth}\right)$. Under the same conditions of Theorem~\ref{thm:estimation},  and $\frac{s\log p\log n\cdot({\sqrt{s}\vee\log^2p})}{\sqrt n}=o(1)$.  Let
		   $T_{\beta, n}(\bth)={\left(T_n(\bth)\right)_{-1,-1}}\in\R^{2p\times 2p}$, then $j\in[p]$, conditional on $X$, we have
		\beq
		\frac{\langle \bm m_j, \frac{1}{\sqrt{n_T}}\sum_{i=1}^{n_T}\gamma_{\bm\theta,i}^{(T)}(y_i-\langle\xx_i,\bbeta_1^*\rangle)\xx_i\rangle}{\sqrt{\bm m_j^\top {(T_{\beta,n}(\hat\bth^{(T)}))_{1,1}}\bm m_j}}\stackrel{d}{\to} N(0,1),
		\eeq
		where $(T_{\beta,n}(\hat\bth^{(T)}))_{1,1}$ is defined in (\ref{T11}).
\end{lemma}


Similar arguments can be applied to the regression coefficient $\bbeta^*_2$. In Algorithm~\ref{alg:hd:inference}, we summarize our proposed method for obtaining the debiased estimators $\widehat{\bm \beta}_1^u$ and $\widehat{\bm \beta}_2^u$. 

\begin{algorithm}
	\caption{De-biasing EM for High-Dimensional MLR}\label{alg:hd:inference}
	\begin{algorithmic}[1]
		\State {\bf Inputs:} $\gamma_{\bm\theta,i}^{(T)}$, $\hat\bbeta_1^{(T)},\hat\bbeta_2^{(T)}$ and $\lambda_n^{(T)}$ from Algorithm~\ref{alg:hd}, $j\in[p]$, tuning parameter $\mu$, and coverage probability $1-\alpha$.
		\State {\bf Precision matrix approximation:}  Let $\hat\omega^{(T)}=\frac{1}{n_T}
	\sum_{i=1}^{n_T}\gamma_{\bm\theta,i}^{(T)}$, and $\tilde\Sigma_{XX}=\frac{1}{n_T}\sum_{i=1}^{n_T}\xx_i\xx_i^\top$ and for $j\in[p]$, let $\tilde{\bm m}_j$ be the solution of 
		\begin{equation*}
		\begin{aligned}
		& \underset{\bm m_j\in \R^p}{\text{minimize}}
		& &\bm m_j^\top\tilde\Sigma_{XX} \bm m_j\\
		& \text{subject to}
		& & ||\tilde\Sigma_{XX} \bm m_j- e^{(p)}_j||_\infty\le \mu;\\
		& & &\|\bm m_j\|_1\le C\sqrt{\log n}.
		\end{aligned}
		\end{equation*}
		 Let $\bm m_j=\tilde{\bm m}_j/{\hat\omega^{(T)}}$.
		\State {\bf De-biasing:} For $\ell=1,2$, let $\widehat{\bm \beta}_{\ell j}^u=\hat\bbeta_{\ell j}^{(T)}+\lambda_n^{(T)} \bm m_j^\top\partial ||\hat\bbeta_\ell^{(T)}||_1$.
		\State {\bf Variance estimation}:
		For $\ell=1,2$, let $$
		\hat v_{\ell j}=\bm m_j^\top \left((T_n(\hat\bth^{(T)}))_{\ell,\ell}\right)\bm m_j,$$
		where $(T_n(\hat\bth))_{\ell,\ell}$ are defined in (\ref{T11}) and (\ref{T22}).
		\State Output $(\widehat{\bm \beta}_{1 j}^u,\hat v_{1j})$ and $(\widehat{\bm \beta}_{2 j}^u,\hat v_{2j})$.
		
	\end{algorithmic}
\end{algorithm}

The following theorem establishes the asymptotic normality of the the individual components of these debiased estimators, which can be directly used for performing hypotheses testing or constructing confidence intervals.
\begin{theorem} \label{EM:CI}
	{Under the same conditions of Theorem~\ref{thm:estimation}.} We further assume that $\|\Sig\|_2,\|\Sig^{-1}\|_1\le L$ for some $L>0$ and $\frac{s\log p\log n\cdot({\sqrt{s}\vee\log^2p})}{\sqrt n}=o(1)$, {and the tuning parameters $\mu=C'\sqrt{\frac{\log p}{n}}\log n$, $C=L$}. Then for any $j\in[p]$ and $\ell=1,2$, conditional on $\xx$, as $n\to\infty$ $$
	\frac{\sqrt{n_T}\left(\widehat{\bm \beta}_{\ell j}^u- \bbeta_{\ell j}^{*}\right)}{\sqrt{\hat v_{\ell j}}}\stackrel{d}{\to} N(0,1).
	$$
\end{theorem}

In MLR, sometimes it is also of interest to know in which features do the associations between $y$ and $\xx$  differ. In other words, one would like to make inference about the differential parameter $(\bbeta_{1j}^*- \bbeta_{2j}^*)$  for some given $j\in [p]$. Towards this end, consider its natural estimator $(\widehat{\bm \beta}_{1j}^u-\widehat{\bm \beta}_{2j}^u)$. It can be shown that a consistent estimator for the variance of $(\widehat{\bm \beta}_{1j}^u-\widehat{\bm \beta}_{2j}^u)$ is given by $$
\tilde{v}_j=\bm m_j^\top \left((T_{\beta,n}(\hat\bth^{(T)}))_{1,1}+(T_{\beta,n}(\hat\bth^{(T)}))_{2,2}-(T_{\beta,n}(\hat\bth^{(T)}))_{1,2}-(T_{\beta,n}(\hat\bth^{(T)}))_{2,1}\right)\bm m_j,$$
where
\begin{align}
&(T_{\beta,n}(\hat\bth^{(T)}))_{1,1}=\frac{1}{n_T}\sum_{i=1}^n\gamma^{(T)}_{\bth,i}\xx_i\xx_i^\top+\frac{2}{n_T}\sum_{i=1}^{n_T}\frac{(y_i-\langle \xx_i,\hat\bbeta_1^{(T)}\rangle )^2}{\eta(\hat\bth^{(T)})}\xx_i\xx_i^\top, \label{T11}\\
&(T_{\beta,n}(\hat\bth^{(T)}))_{2,2}=\frac{1}{n_T}\sum_{i=1}^n(1-\gamma^{(T)}_{\bth,i})\xx_i\xx_i^\top+\frac{2}{n_T}\sum_{i=1}^n\frac{(y_i-\langle \xx_i,\hat\bbeta_2^{(T)}\rangle )^2}{\eta(\hat\bth^{(T)})}\xx_i\xx_i^\top, \label{T22}\\
&(T_{\beta,n}(\hat\bth^{(T)}))_{2,1}=(T_{\beta,n}(\hat\bth^{(T)}))_{1,2}=\frac{2}{n_T}\sum_{i=1}^n\frac{{(\langle\xx_i,\hat\bbeta_1^{(T)}\rangle-y_i)}\cdot(y_i-\langle \xx_i,\hat\bbeta_2^{(T)}\rangle )}{\eta(\hat\bth^{(T)})}\xx_i\xx_i^\top, \label{T12}
\end{align}
and \begin{align} \label{bth}
\eta(\hat\bth^{(T)})=&\sigma^2\bigg[\hat{\omega}^{(T)}+(1-\hat{\omega}^{(T)})\exp\bigg\{\frac{(2y_i-\langle\xx_i,\hat\bbeta_1^{(T)}+\hat\bbeta_2^{(T)}\rangle)\cdot\langle\xx_i,\hat\bbeta_2^{(T)}-\hat\bbeta_1^{(T)}\rangle}{2\sigma^2}\bigg\} \bigg] \nonumber \\
&\times \bigg[1-\hat{\omega}^{(T)}+\hat{\omega}^{(T)}\exp\bigg\{-\frac{(2y_i-\langle\xx_i,\hat\bbeta_1^{(T)}+\hat\bbeta_2^{(T)}\rangle)\cdot\langle\xx_i,\hat\bbeta_2^{(T)}-\hat\bbeta_1^{(T)}\rangle}{2\sigma^2}\bigg\}\bigg].
\end{align}
Similar toTheorem~\ref{EM:CI}, the following theorem establishes the asymptotic normality of the estimator $(\widehat{\bm \beta}_{1j}^u-\widehat{\bm \beta}_{2j}^u)$.

\begin{theorem}\label{thm:inference} 
	{Under the same conditions of Theorem~\ref{thm:estimation}.} We further assume that $\|\Sigma^{-1}\|_1\le L$ for some $L>0$ and $\frac{s\log p\log n}{\sqrt n}\to 0$. Then for any $j\in[p]$, as $n\to\infty$, conditional on  $\xx$, $$
	\frac{\sqrt{n_T}\left((\widehat{\bm \beta}_{1j}^u-\widehat{\bm \beta}_{2j}^u)- (\bbeta_{1j}^{*}-\bbeta_{2j}^{*})\right)}{\sqrt{\tilde{v}_j}}\stackrel{d}{\to} N(0,1).
	$$
\end{theorem} 

\subsection{Asymptotic Confidence Intervals}

Given the asymptotic normality established in Theorems \ref{EM:CI} and \ref{thm:inference}, we are now ready to present the confidence intervals for the individual coordinates $\beta_{lj}$'s for $j\in[p]$ and $l=1,2$ and the differential parameters $(\bbeta_{1j}^*- \bbeta_{2j}^*)$  for  $j\in [p]$. Specifically, let $$
I_{lj}^{(ind)}=[\widehat{\bm \beta}_{\ell j}^u-z_{\alpha/2}\sqrt{\hat v_{\ell j}}, \;\;\widehat{\bm \beta}_{\ell j}^u+z_{\alpha/2}\sqrt{\hat v_{\ell j}}], \quad\text{ for $j\in[p]$ and $l=1,2$},
$$
and $$
I_{j}^{(dif)}=[(\widehat{\bm \beta}_{1j}^u-\widehat{\bm \beta}_{2j}^u)-z_{\alpha/2}\sqrt{\tilde{v}_j}, (\widehat{\bm \beta}_{1j}^u-\widehat{\bm \beta}_{2j}^u)+z_{\alpha/2}\sqrt{\tilde{v}_j}], \quad\text{ for $j\in[p]$},
$$
where $z_{\alpha/2}$ is the $\alpha/2$-th quantile of a standard normal distribution.  The following theorem provides the asymptotic guarantee for the validity of these confidence intervals.
\begin{theorem}\label{thm:CI}
Under the conditions of Theorem~\ref{EM:CI}, the confidence intervals $I_{lj}^{(ind)}$ and $I_{j}^{(dif)}$ are asymptotically valid, that is,
\begin{align*}
&\lim_{n\to\infty}\Pro(\bm\beta_{lj}\in I_{lj}^{(ind)})=1-\alpha, \quad\text{ for $j\in[p]$ and $l=1,2$};\\
&\lim_{n\to\infty}\Pro(\bm\beta_{1j}-\bm\beta_{2j}\in I_{j}^{(dif)})=1-\alpha, \quad\text{ for $j\in[p]$}.
\end{align*}
\end{theorem}


\section{LARGE-SCALE MULTIPLE TESTING}

\subsection{The Multiple Testing Procedure}

In this section, we consider simultaneous testing of the following null hypotheses
\[
H_{0j}: \bbeta_{1j}^*=\bbeta_{2j}^*=0,\quad 1\le j\le p.
\]
Apart from identifying as many nonzero coordinates as possible, to obtain results of practical interest, we
would also like to control the false discovery rate (FDR) as well as the false discovery proportion (FDP).

Specifically, since each individual hypothesis $H_{0j}$ is a composite of two hypotheses with $H_{0j} = H^{(1)}_{0j}\cap  H^{(2)}_{0,j}$ where $H^{(\ell)}_{0j} :  \bbeta_{\ell j}^*=0,$ we can construct standardized statistics
\[
T_j^{(\ell)} = \frac{\hat{\bbeta}_{\ell j}^u}{\hat{v}_{\ell j}/\sqrt{n}},\quad\text{for $j=1,...,p$ and $\ell=1,2$}.
\]
For a given threshold level $t>0$, each individual partial hypothesis $H^{(\ell)}_{0j}:\bbeta_{\ell j}^*=0$ is rejected if $|T^{(\ell)}_j|\ge t$. Hence if we propose a test statistic
\[
T_j = \max \{|T_j^{(1)}|,|T_j^{(2)}|\} 
\]
for each null hypothesis $H_{0j}$ and we reject $H_{0j}$ whenever $T_j \ge t$, then for each $t$, we can define 
\[
\text{FDP}(t) = \frac{\sum_{j\in \mathcal{H}_0} I\{ T_j \ge t\}}{\max \big\{\sum_{j=1}^p I\{ T_j \ge t\},1   \big\}}, \quad\quad \text{FDR}(t) = \E [\text{FDP}(t)].
\]
In order to control the FDR/FDP at a pre-specified level $0<\alpha <1$, we can set the threshold level as
\beq \label{ideal.fdr}
\tilde{t}_1 = \inf\bigg\{ 0\le t\le b_p: \frac{\sum_{j\in \mathcal{H}_0} I\{ T_j \ge t\}}{\max \big\{\sum_{j=1}^p I\{ T_j \ge t\},1   \big\}} \le \alpha  \bigg\}
\eeq
for some $b_p$ to be determined later.
In general, the ideal choice $\tilde{t}_1$ is unknown since it depends on the knowledge of the true null $\mathcal{H}_0$. Inspired by the Gaussian approximation idea proposed by \cite{liu2013gaussian}, we first substitute the numerator in (\ref{ideal.fdr}) by its upper bound
\[
\sum_{j\in \mathcal{H}_0} I\{ T_j \ge t\} \le \sum_{j\in \mathcal{H}_0} I\{ |T^{(1)}_j| \ge t\}+\sum_{j\in \mathcal{H}_0} I\{ |T^{(2)}_j| \ge t\},
\]
and then use Gaussian tails to approximate the counts $\sum_{j\in \mathcal{H}_0} I\{ |T^{(\ell)}_j| \ge t\}$ for $\ell=1,2$. Specifically, let $G^{(\ell)}_0(t)$ be an estimate of the proportion of the nulls falsely rejected by the test $I\{|T_j^{(\ell)}|\ge t\}$ among all the true nulls at the threshold level $t$, so that
\beq
G^{(\ell)}_0(t) = \frac{1}{ |\mathcal{H}_0|}\sum_{j\in \mathcal{H}_0}I\{ |T^{(\ell)}_j| \ge t\},\quad\quad \ell=1,2.
\eeq
Let $G(t) = 2-2\Phi(t)$ be the tails of normal distribution. We will show that, asymptotically, we can use $G(t)$ to approximate $G_0^{(\ell)}(t)$ for $\ell=1,2$.
Therefore, we have the following multiple testing procedure controlling the FDR and the FDP.

\begin{proc}
	Let $0<\alpha <1$, $b_p=\sqrt{2\log p-2\log\log p}$ and define
	\beq \label{t.hat.fdr}
	\hat{t} = \inf\bigg\{ 0\le t\le b_p: \frac{pG(t)}{\max \big\{\sum_{j=1}^p I\{ |T_j| \ge t\},1   \big\}} \le \alpha /2 \bigg\}.
	\eeq
	If $\hat{t}$ in (\ref{t.hat.fdr}) does not exist, then let $\hat{t} = \sqrt{2\log p}$. We reject $H_{0,j}$ whenever $|T_j| \ge \hat{t}$.
\end{proc}

\subsection{Theoretical Properties}

For $\ell=1,2$, let $\Gamma_\ell=M^\top (\E_{\bth^*}[T_n(\bth^*)])_{\ell,\ell} M$ where $M$ has its $j$-th column as $\bm m_j$ and let $D_\ell$ be the diagonal of $\Gamma_\ell$. We define $D_\ell^{-1/2}\Gamma_\ell D_\ell^{-1/2}=(\rho_{jk}^{(\ell)})_{1\le j,k\le p} $ and denote $\mathcal{B}_\ell(\delta) = \{ (j,k): |\rho^{(\ell)}_{jk}| \ge \delta, i \ne j  \}$ and $\mathcal{A}_\ell(\epsilon) = \mathcal{B}_\ell((\log p)^{-2-\epsilon}).$\\
\indent ({A3}). Suppose that for any $\ell=1,2$, there is some $\epsilon >0$ and $q>0$, such that
\[
\sum_{j,k\in \HH_0:(j,k)\in \mathcal{A}_\ell(\epsilon)} p^{\frac{2|\rho^{(\ell)}_{jk}|}{1+|\rho^{(\ell)}_{jk}|}+q} = O(p^2/(\log p)^2).
\]
The following theorem shows the asymptotic control of FDR and FDP of our procedure.
\bet \label{fdr}
Under the conditions of Theorem 2, if (A3) holds, then for $\hat{t}$ defined in Procedure 1, 
\beq
\lim_{(n,p)\rightarrow \infty} \frac{\textup{FDR}(\hat{t})}{\alpha p_0/p} \le 1, \quad\quad \lim_{(n,p)\rightarrow \infty} \Pro\bigg( \frac{\textup{FDP}(\hat{t})}{\alpha p_0/p}\le 1+\epsilon \bigg) = 1
\eeq
for any $\epsilon>0$. 
\eet

\section{SIMULATION STUDIES}\label{sec:sim}

In this section, we evaluate the numerical performance of the proposed methods. For both estimation and large-scale multiple testing, the empirical results in various settings demonstrate the numerical advantages of the proposed procedures over alternative methods.

\subsection{Estimation}

For estimation, we let the dimension of the covariates $p$ range from 600 to 1000, the sparsity $s$ vary from 10 to 30, and set the sample size $n=400$. We also set the mixture proportion $\omega^*=0.3$ and the noise level $\sigma^2 =1$. 
The design covariates $\xx_i$'s are generated from a multivariate Gaussian distribution with covariance matrix $\Sig=\Sig_M$, where $\Sig_M$ is a $p\times p$ blockwise diagonal matrix of $10$ identical unit diagonal Toeplitz matrices whose off-diagonal entries descend from 0.4 to 0 (see Supplementary Material for the explicit form).
For the two regression coefficients $\bbeta^*_1$ and $\bbeta^*_2$, for some fixed $\rho>0$, we set $\bbeta^*_{1j}=\rho\cdot 1\{1\le j\le s\}$ and  $\bbeta^*_{2j}=-\rho\cdot 1\{p/2+1\le j\le p/2+s\}$ so that each of the coefficient vectors is $s$-sparse. 

In particular, for our proposed methods, as of practical interest, we assume that the noise level $\sigma^2$ is unknown and also needs to be estimated at each iteration. Specifically, for any $t\ge 1$, in the M-Step of Algorithm \ref{alg:hd}, we define 
\[
(\hat{\sigma}^2_1)^{(t)}={\frac{1}{n}\sum_{i=1}^n \gamma_{\bth,i}^{(t)}(y_i-\langle \xx_i,\bbeta\rangle )^2},\qquad (\hat{\sigma}^2_2)^{(t)}={\frac{1}{n}\sum_{i=1}^n (1-\gamma_{\bth,i}^{(t)})(y_i-\langle \xx_i,\bbeta\rangle )^2},
\]
and set $(\hat{\sigma}^{2})^{(t)}={[(\hat{\sigma}^2_1)^{(t)}+(\hat{\sigma}^2_2)^{(t)}]/2}$. The variance estimator $(\hat{\sigma}^2)^{(t)}$ is then used as a substitute for $\sigma^2$ in the subsequent E-Step.
Throughout, we set $T=30$, $\kappa=0.3$ and $C=0.8$. 

We consider two initializations for our proposed algorithm. We start with fitting a Lasso to the mixed samples, which results to a coarse but useful variable screening. Combining the response variable $y$ and the Lasso selected covariates, we use one of the following high-dimensional clustering methods to divide the samples. In particular, our two initialization corresponds to the \texttt{emgm} function in the R package \texttt{xLLiM}, and the \texttt{hddc} algorithm in the R package \texttt{HDclassif}. Once we obtain an initial two-group clustering of samples, we can fit the Elastic Net separately using the samples within each group. The resulting regression coefficients will be used as initial values $\hat{\bbeta}_1^{(0)}$ and $\hat{\bbeta}_2^{(0)}$, respectively. For the above Elastic Net algorithm \citep{zou2005regularization}, we set the elasticnet mixing parameter as $0.5$.

We evaluate and compare the empirical performances of 1) GLLiM: the Gaussian Locally Linear Mapping EM algorithm proposed by \cite{deleforge2015high}, which is implemented by the \texttt{gllim} function in the R package \texttt{xLLiM}; 2)  Initial1: fit Elastic Net separately to the clusters determined by Lasso+\texttt{emgm}; 3) Initial2: fit Elastic Net separately to the clusters determined by Lasso+\texttt{hddc}; 4) MIREM1 : our proposed algorithm based on initialization Initial1; and 5) MIREM2: our proposed algorithm based on initialization Initial2.

The estimation performance is evaluated using the empirical mean-squared error (EMSE): for $N$ rounds of simulations and estimators $(\hat{\bbeta}_1^r,\hat{\bbeta}_2^r)$ obtained in the $r$-th round, we define
\[
\text{EMSE} = \min\bigg\{\frac{1}{N}\sum_{r=1}^N [\|\hat{\bbeta}_1^r-\bbeta_1^*\|_2+\|\hat{\bbeta}_2^r-\bbeta_2^*\|_2],\frac{1}{N}\sum_{r=1}^N[ \|\hat{\bbeta}_1^r-\bbeta_2^*\|_2+\|\hat{\bbeta}_2^r-\bbeta_1^*\|_2] \bigg\}.
\]


\begin{table}[h!]
	\centering
	\caption{Comparison of empirical mean-squared error (EMSE) of different methods with $\omega^*=0.3$ and $n=400$}
	\begin{tabular}{cccccc|ccccc}
		\hline
		 &\multicolumn{5}{c|}{$\rho=0.45$} & \multicolumn{5}{c}{$\rho=0.85$}\\
		 $p=$&600& 700 & 800 & 900 & 1000 & 600& 700 & 800 & 900 & 1000 \\
		\hline
		&\multicolumn{9}{c}{$s=10$}&\\
		GLLiM &    2.86 & 2.95 &  3.17 & 3.20 & 3.26& 5.09 & 5.11 & 5.12  & 5.13 &  5.16  \\
		Initial1 &   3.19 & 3.00 & 3.12 & 3.12   & 3.27 & 6.07 & 5.94& 6.00& 6.06& 6.27\\
		Initial2 &  2.43 & 2.43 & 2.50 & 2.41   & 2.59 & 5.08 & 5.20&5.22&5.06&4.76 \\
		MIREM1 &  1.40  & 1.40 & 1.42 & 1.42   & 1.43 & 1.18 & 1.18& 1.18&1.21 & 1.23\\
		MIREM2 &  1.73 & 1.81 & 1.75 & 1.79   & 1.81 & 2.85 & 2.17&2.29&2.61 &2.66 \\
		\hline
		&\multicolumn{9}{c}{$s=15$}&\\
		GLLiM &    3.16  & 3.21 &  3.27 & 3.30 & 3.34& 6.26 & 6.29& 6.35&6.31 & 6.37\\
Initial1 &   4.19 & 4.21 & 4.21  & 4.21  & 4.40 & 9.04 & 9.02 &9.01&8.99 &8.97 \\
Initial2 &  3.21 & 3.15 & 3.25& 3.16  & 3.08 & 8.68 & 7.66&8.35& 8.00 & 7.71\\
MIREM1 &   1.42 & 1.47 & 1.45 & 1.47   & 1.49 & 1.26 & 1.31&1.62 &1.57 & 1.56 \\
MIREM2 &  1.92 & 1.98 & 2.04&  2.02  & 1.94 & 3.36 & 3.65&3.48&  3.75 & 4.03\\
		\hline
		&\multicolumn{9}{c}{$s=20$}&\\
		GLLiM &    3.63  & 3.66 &  3.66 & 3.68 & 3.73 & 7.31 & 7.32& 7.34&7.32 & 7.35 \\
Initial1 &   5.45 & 5.38 & 5.68 & 5.62  & 5.52 & 12.18 & 12.11& 12.09&12.15 &11.89 \\
Initial2 &  4.35 & 3.92 & 4.11 & 4.39  & 4.18 & 11.80 & 10.39&10.61& 10.81& 10.59\\
MIREM1 &   1.57 & 1.58 & 1.60 & 1.56   & 1.61 & 1.77 & 2.43 &2.15& 1.77 &2.14\\
MIREM2 &  2.49 & 2.45 & 2.32 & 2.34   & 2.41 & 5.14 & 4.91&5.19&5.23 &5.31\\
		\hline
		&\multicolumn{9}{c}{$s=25$}&\\
		GLLiM &   4.08  & 4.08 &  4.14 & 4.11 & 4.14& 8.23 & 8.23& 8.24&8.24&8.24\\
Initial1 &   6.96 & 7.04 & 6.91 & 6.99  & 6.88 & 15.53 & 15.29 &15.22&15.08&15.24\\
Initial2 &  5.68 & 6.02 & 5.74 & 5.57   & 5.78 & 13.74 & 13.85& 13.86&13.48&13.35 \\
MIREM1 &   1.82 & 1.71 & 1.73 & 1.74   & 1.75 & 3.60 & 4.41&  3.93& 4.81&5.70  \\
MIREM2 &  3.00 & 2.88 & 2.84 & 3.10  & 2.99 & 7.98 & 7.26&  6.71& 7.40 & 7.02\\
		\hline
		&\multicolumn{9}{c}{$s=30$}&\\
		GLLiM &    4.51  & 4.52 &  4.53 & 4.58 & 4.56&9.06 & 9.07& 9.04&9.05 &9.08 \\
Initial1 &   8.35  & 8.30 & 8.55 & 8.59   & 8.55 & 18.63 & 18.46& 18.17&18.35 &17.77\\
Initial2 &  7.61  & 7.30 & 7.50 & 7.35   & 7.75 & 16.41 & 17.10& 16.79&16.24 &15.44 \\
MIREM1 &  1.99 & 1.94 & 1.95 & 1.91   & 1.95 & 5.36 & 8.34& 6.76&10.50 &8.79\\
MIREM2 &  3.64 & 3.50 & 3.48 & 3.68   & 3.89 & 8.32 & 9.26&9.65&9.66&9.00 \\
		\hline
	\end{tabular}
	\label{table:t1}
\end{table}

In Table \ref{table:t1}, we show the EMSEs calculated from $N=500$ rounds of simulations. We observe that both MIREM1 and MIREM2 outperform the other three methods across almost all the settings. As dimension $p$, the sparsity $s$, or the signal magnitude $\rho$ increases, all the methods show increased estimation errors. In addition, comparing our proposed methods MIREM1 and MIREM2, we find that MIREM1 has better performance than MIREM2 in almost all the settings. The EM based GLLiM method, with 100 iterations, performs slightly better than our initializations, but our proposed MIREM algorithms, with only $T=30$ iterations, have superior performance, suggesting significant improvement upon the initial estimators.

\subsection{Large-scale Multiple Testing and FDR Control}

In this section, the empirical performance of our proposed multiple testing procedure is evaluated under different settings. Specifically, we vary the number of covariates $p$ from 800 to 1000, the sparsity level $s$ from 10, 15 to 20, and set the sample size $n$ as 300 or 400. The two regression coefficients  $\bbeta^*_1$ and $\bbeta^*_2$, the design covariates, the mixing proportion $\omega^*$ and the number of  iterations $T$ are the same as previous simulations with $\rho=0.45$. About our proposed method, in light of the results from the previous section, we will focus on MIREM1 instead of MIREM2 for its superior performance across most settings.
To the best of our knowledge, there is no existing method for multiple testing in mixed linear regression models. So we compare the empirical FDRs and powers of our proposed testing procedure to the Benjamini-Yekutieli (B-Y) procedure \citep{benjamini2001control} applied to our proposed test statistics for individual tests. To illustrate the necessity of fitting a mixed linear regression model when the underlying model is indeed a mixture, we also evaluate the performance of the multiple testing procedure based on ordinary debiased Lasso estimators \citep{javanmard2019false}, denoted as dLasso,  designed for the linear regression models.


\begin{table}[h!]
	\centering
	\caption{Empirical powers and FDRs with $\alpha=0.1$, $\omega^*=0.3$ and $n=300$}
	\begin{tabular}{cccccc|ccccc}
		\hline
		&\multicolumn{5}{c|}{Powers} & \multicolumn{5}{c}{FDRs}\\
		$p=$& 800& 850 & 900 & 950 & 1000 &800& 850 & 900 & 950 & 1000 \\
		\hline
		&\multicolumn{9}{c}{$s=10$}&\\
		MIREM1 &  0.459  & 0.459 & 0.430 & 0.465   & 0.441 & 0.009 & 0.009 & 0.014&0.012 & 0.025 \\
		B-Y &  0.344 & 0.344 & 0.290 & 0.346   & 0.332 & $<$0.001 & $<$0.001&$<$0.001&$<$0.001 & $<$0.001 \\
		dLasso &  0.934 & 0.934 & 0.930 & 0.918  & 0.896 & 0.958 &0.958&0.960 &0.963 &0.965 \\
		\hline
		&\multicolumn{9}{c}{$s=15$}&\\
		MIREM1 &  0.582 & 0.592 & 0.563 & 0.623   & 0.609 & 0.022 & 0.024& 0.028&0.028 & 0.030 \\
		B-Y &  0.510& 0.530 & 0.484 & 0.550  & 0.551 & $<$0.001 & $<$0.001& $<$0.001 &$<$0.001  &$<$0.001  \\
		dLasso &  0.897 & 0.922 & 0.916 & 0.914   & 0.901 & 0.946& 0.948 &0.951 &0.953  &0.956 \\
		\hline
		&\multicolumn{9}{c}{$s=20$}&\\
		MIREM1 &  0.724  & 0.744 &0.723 & 0.756 & 0.778 & 0.088 & 0.086 & 0.071 &0.089  & 0.110 \\
		B-Y &  0.621 & 0.635 & 0.617 & 0.657   & 0.672 & 0.001 & 0.001&0.001&0.001 &0.001\\
		dLasso &  0.882 & 0.909 & 0.897 & 0.894    & 0.896 & 0.936 & 0.937&0.942 &0.945 &0.947 \\
		\hline
	\end{tabular}
	\label{table:t2}
\end{table}

\begin{table}[h!]
	\centering
	\caption{Empirical powers and FDRs with $\alpha=0.1$, $\omega^*=0.3$ and $n=400$}
	\begin{tabular}{cccccc|ccccc}
		\hline
		&\multicolumn{5}{c|}{Powers} & \multicolumn{5}{c}{FDRs}\\
		$p=$& 800& 850 & 900 & 950 & 1000 &800& 850 & 900 & 950 & 1000 \\
		\hline
		&\multicolumn{9}{c}{$s=10$}&\\
		MIREM1 &  0.864  & 0.805 & 0.774 & 0.796   & 0.846 & 0.046 & 0.017 & 0.036&0.015 & 0.066\\
		B-Y &  0.849 & 0.779 & 0.748& 0.768   & 0.833& 0.001 & $<$0.001&$<$0.001&$<$0.001 & 0.001 \\
		dLasso &  0.977 & 0.973 & 0.975& 0.985  & 0.994 & 0.943 &0.957&0.957&0.961 &0.962 \\
		\hline
		&\multicolumn{9}{c}{$s=15$}&\\
		MIREM1 &  0.847 & 0.863 & 0.859 & 0.859   & 0.877 & 0.044 & 0.040& 0.028&0.052 & 0.044 \\
	B-Y &  0.825& 0.842 & 0.843 & 0.834  & 0.857 & 0.001 & 0.001&0.001 &0.001  &0.001  \\
	dLasso &  0.964 & 0.968 & 0.968 & 0.965   & 0.973& 0.945& 0.949 &0.952 &0.954  &0.956 \\
		\hline
		&\multicolumn{9}{c}{$s=20$}&\\
		MIREM1 &  0.933  & 0.914 &0.935 & 0.911 & 0.920 & 0.105 & 0.125 & 0.109 &0.104  & 0.112 \\
	B-Y &  0.905 & 0.873 & 0.900 & 0.877   & 0.889 & 0.001 & 0.002&0.001&0.001 &0.001\\
	dLasso &  0.983 & 0.969 & 0.969 & 0.974    & 0.975 & 0.932 & 0.941&0.942 &0.947 &0.947 \\
		\hline
	\end{tabular}
	\label{table:t3}
\end{table}

From Tables \ref{table:t2}  and \ref{table:t3}, we find that the B-Y procedure and our proposed multiple testing procedure are both able to control the FDR below or around the nominal level $\alpha=0.1$, whereas the dLasso fails to control the FDR, as a consequence of its inability to capture the mixture structure. In particular, the empirical FDRs of our proposed test procedure are closer to the nominal level $\alpha$ in comparison to the rather conservative B-Y procedure, yielding improved empirical powers of our proposed method across all the settings. In particular, by inspecting the intermediate steps of the dLasso method, we found that, due to the failure to account for the mixture components, dLasso significantly underestimates the standard errors of the debiased Lasso estimators and therefore the individual $p$-values, which explains the anticonservativeness of the dLasso method.

\section{ANALYSIS OF A MULTIPLEX IMAGE CYTOMETRY DATASET}

\begin{figure}[h!]
	\centering
	\includegraphics[angle=0,width=5cm]{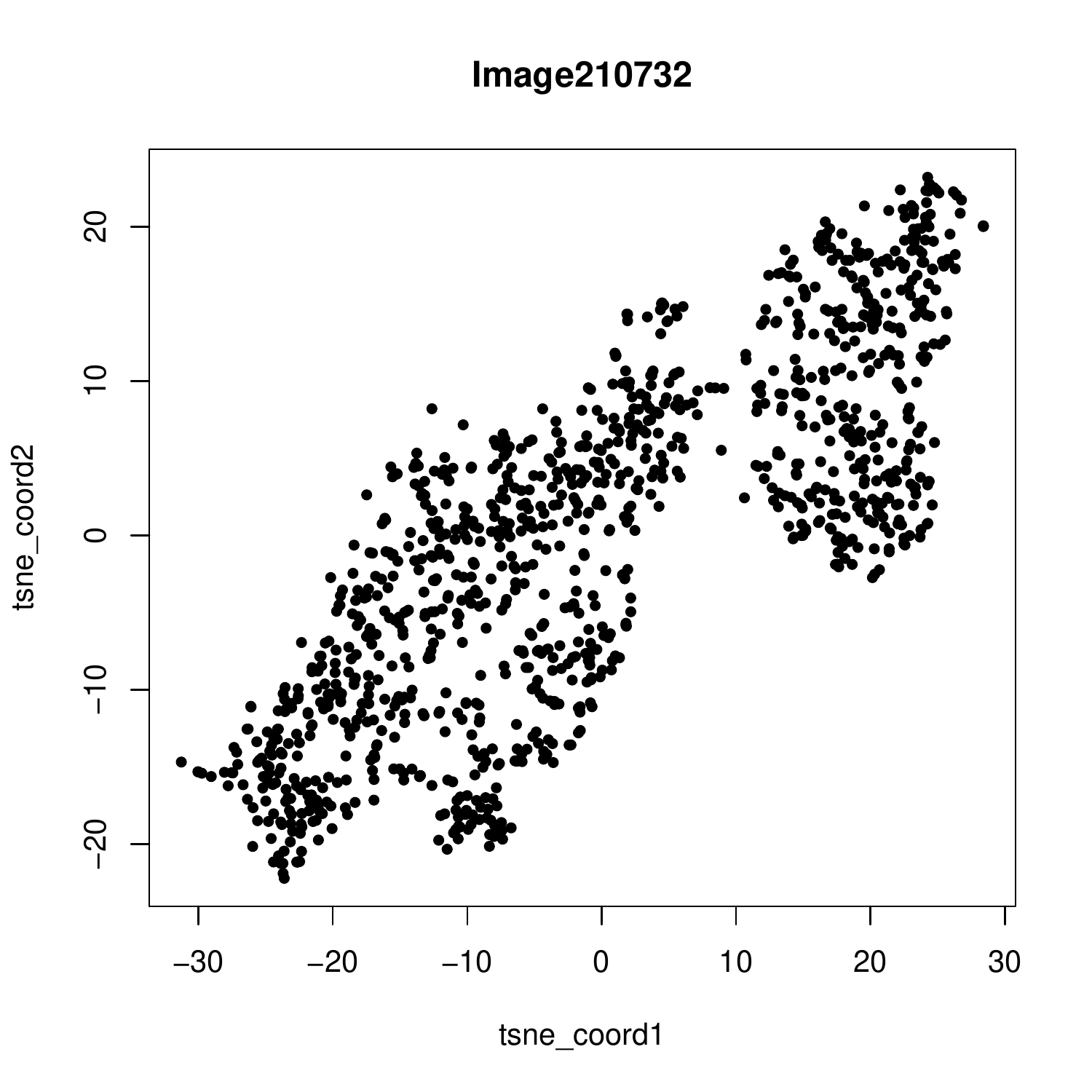}
	\includegraphics[angle=0,width=5cm]{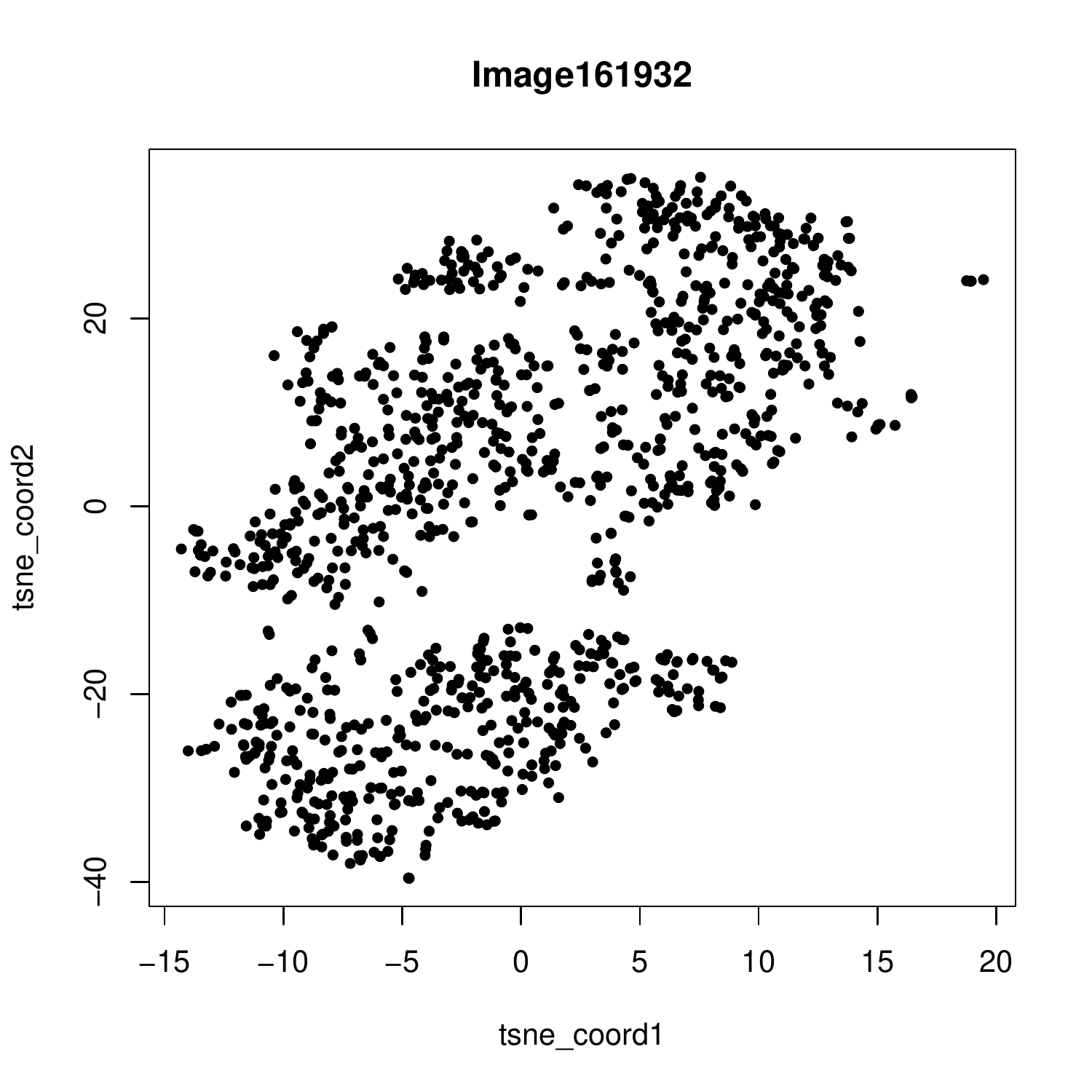}
	\includegraphics[angle=0,width=5cm]{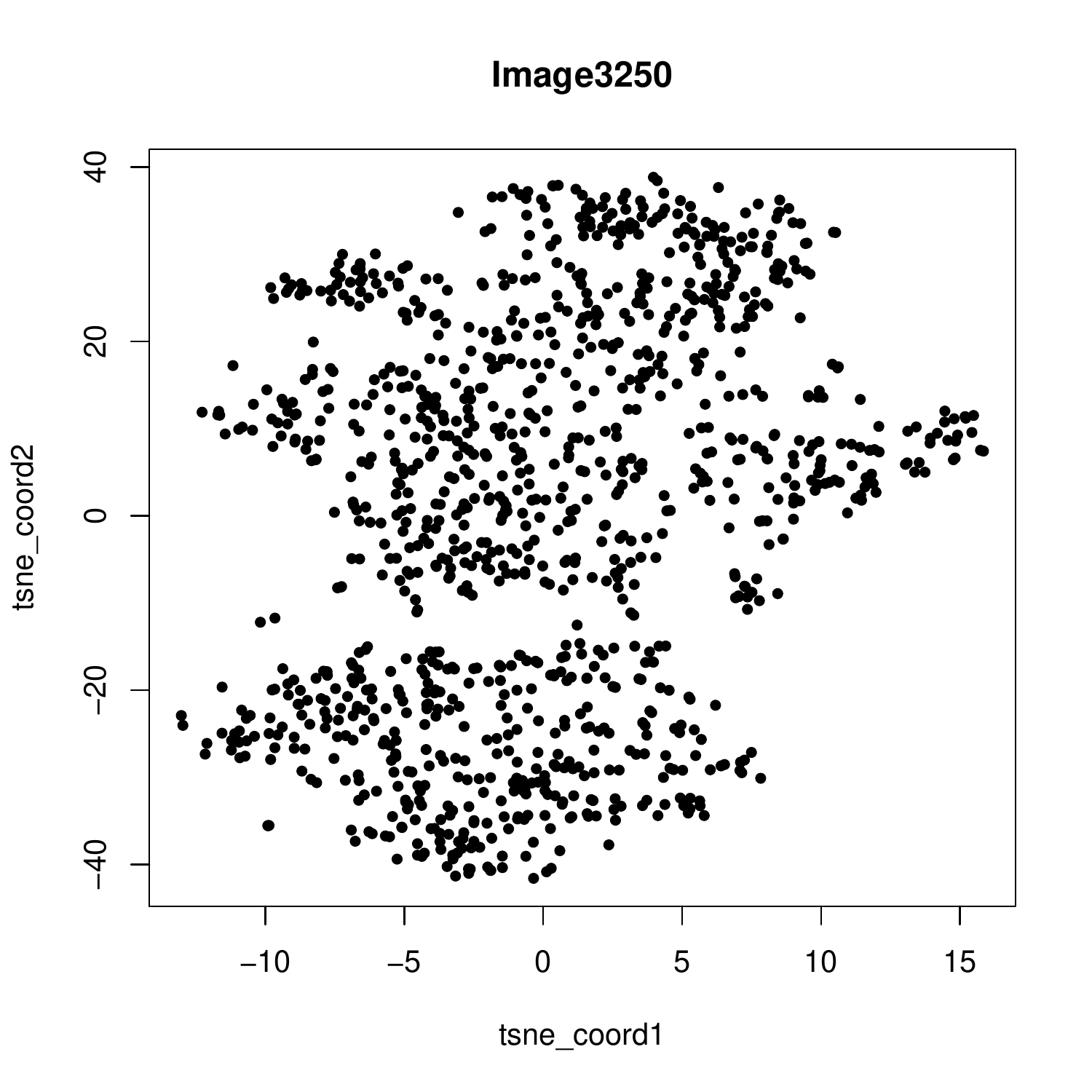}\\
	\includegraphics[angle=0,width=5cm]{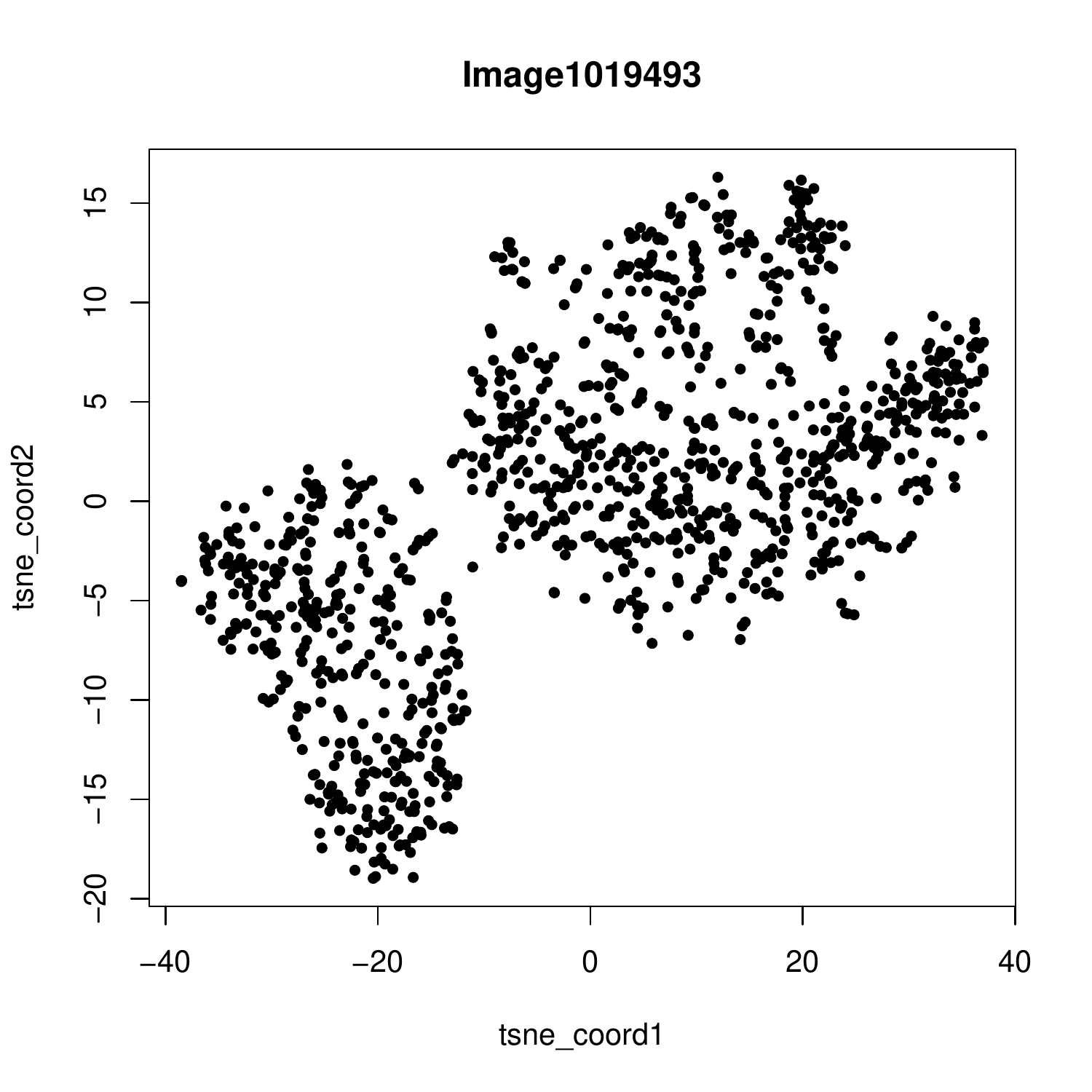}
	\includegraphics[angle=0,width=5cm]{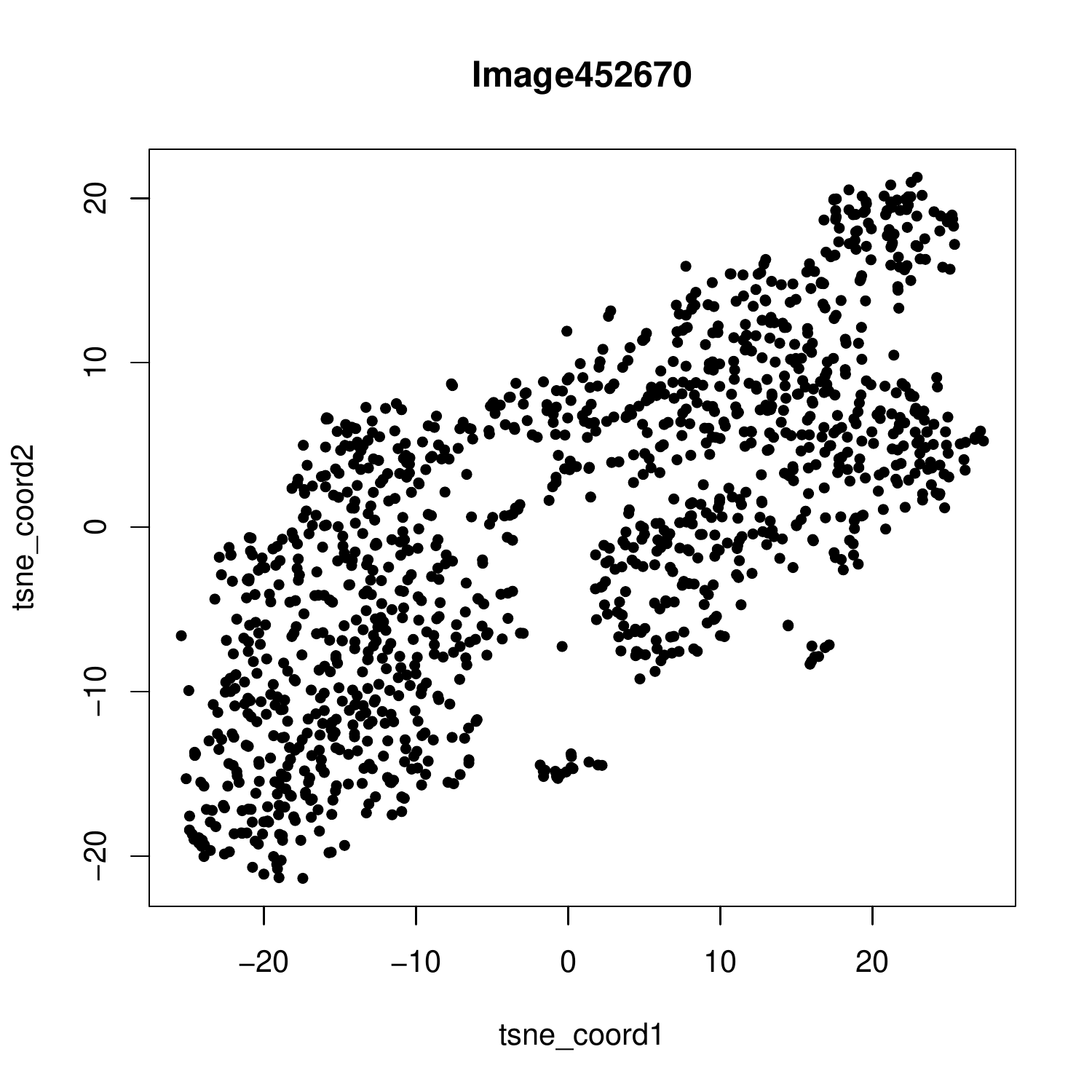}
	\includegraphics[angle=0,width=5cm]{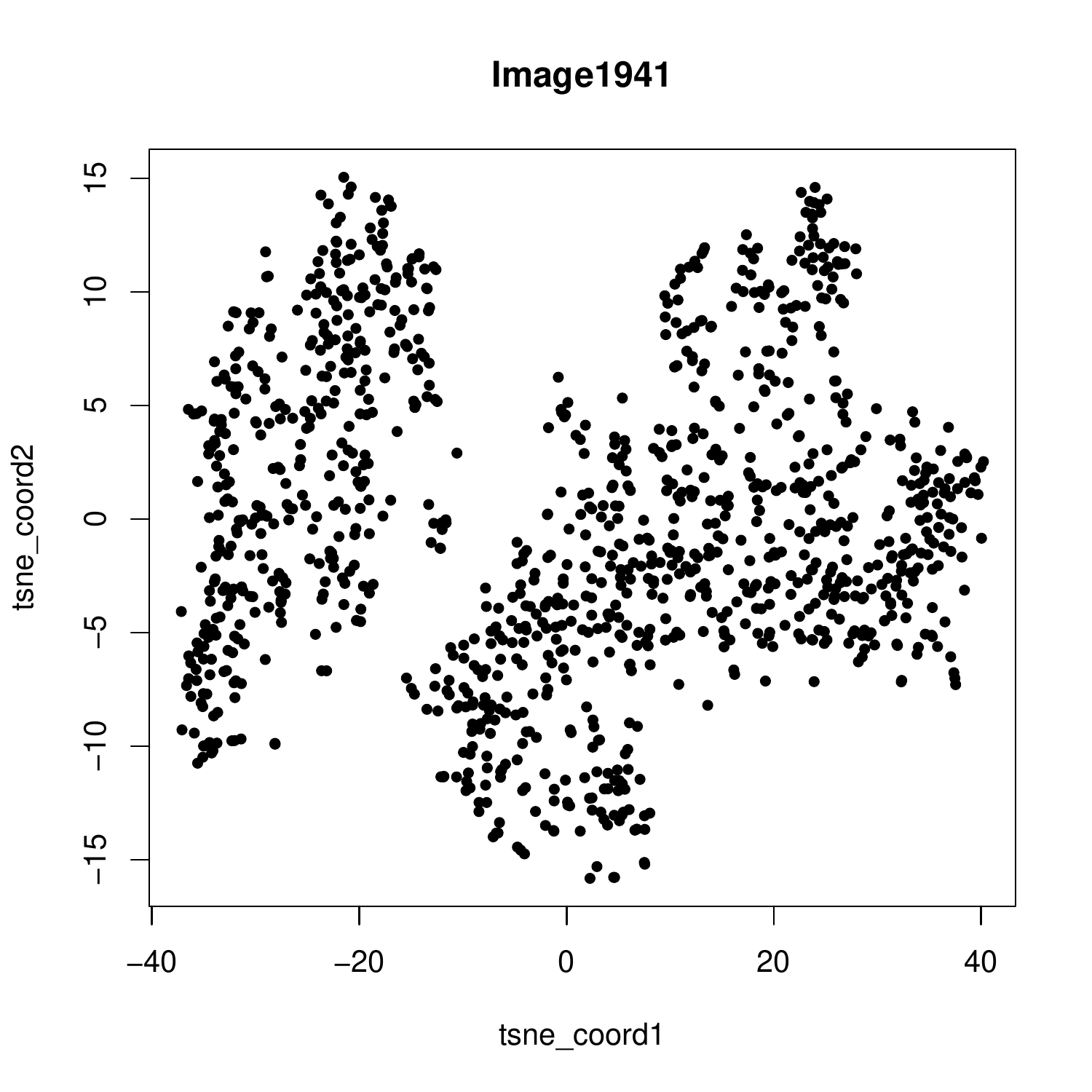}
	\caption{tSNE plots of cells in six randomly selected images/samples based on the expression levels of 30 epitopes or protein  markers. } \label{tsne}
\end{figure}

In this section, we apply our proposed methods to analyze a multiplex image cytometry dataset studied by \cite{schapiro2017histocat}. Specifically, our dataset contains cellular phenotypes visualized by the imaging mass cytometry (IMC). By pairing classic immunohistochemistry staining, high-resolution tissue laser ablation, and mass cytometry, IMC can measure abundances of more than 40 unique metal-isotope-labeled tissue-bound antibodies simultaneously at a resolution comparable to that of fluorescence microscopy. \cite{schapiro2017histocat} analyzed images collected from 49 diverse breast cancer samples and 3 matched normal tissues, with each image containing cells whose number varies from 266 to 1,454. Among the 30 cellular phenotypes, there are expression levels of 20 different epitopes (e.g., vimentin; and CD68) or combinations of markers (e.g., proliferative Ki-67+ and phospho-S6+). An initial analysis of our image cytometry datasets using tSNE indicates strong evidences of population heterogeneity among the cells within each of the images (see Figure \ref{tsne} for some examples).

We focus on analyzing the conditional dependence network among these 30 protein epitopes and markers based on the single cell data for each of the samples or images.
 It is well known that the conditional dependence network can be modeled by Gaussian graphical model, which can be obtained using node-wise regression \citep{meinshausen2006high,yuan2007model}. In other words, to obtain the dependence between two variables $X$ and $Y$ conditioning on all the other variables $(Z_1,...,Z_d)$, it suffices to perform a linear regression between $(X,Z_1,...,Z_d)$ against $Y$ and assess the coefficient of $X$. The construction of the conditional dependence network thus requires fitting such linear regressions over all the possible variable configurations. However, in our image cytometry data, heterogeneity among different cell-types may induce a mixture of different dependence structures. To address such an issue, we apply our proposed methods based on the sparse mixed linear regression model instead of the ordinary sparse linear regression model for network construction.

\begin{figure}[h!]
	\centering
	\includegraphics[angle=0,width=5cm]{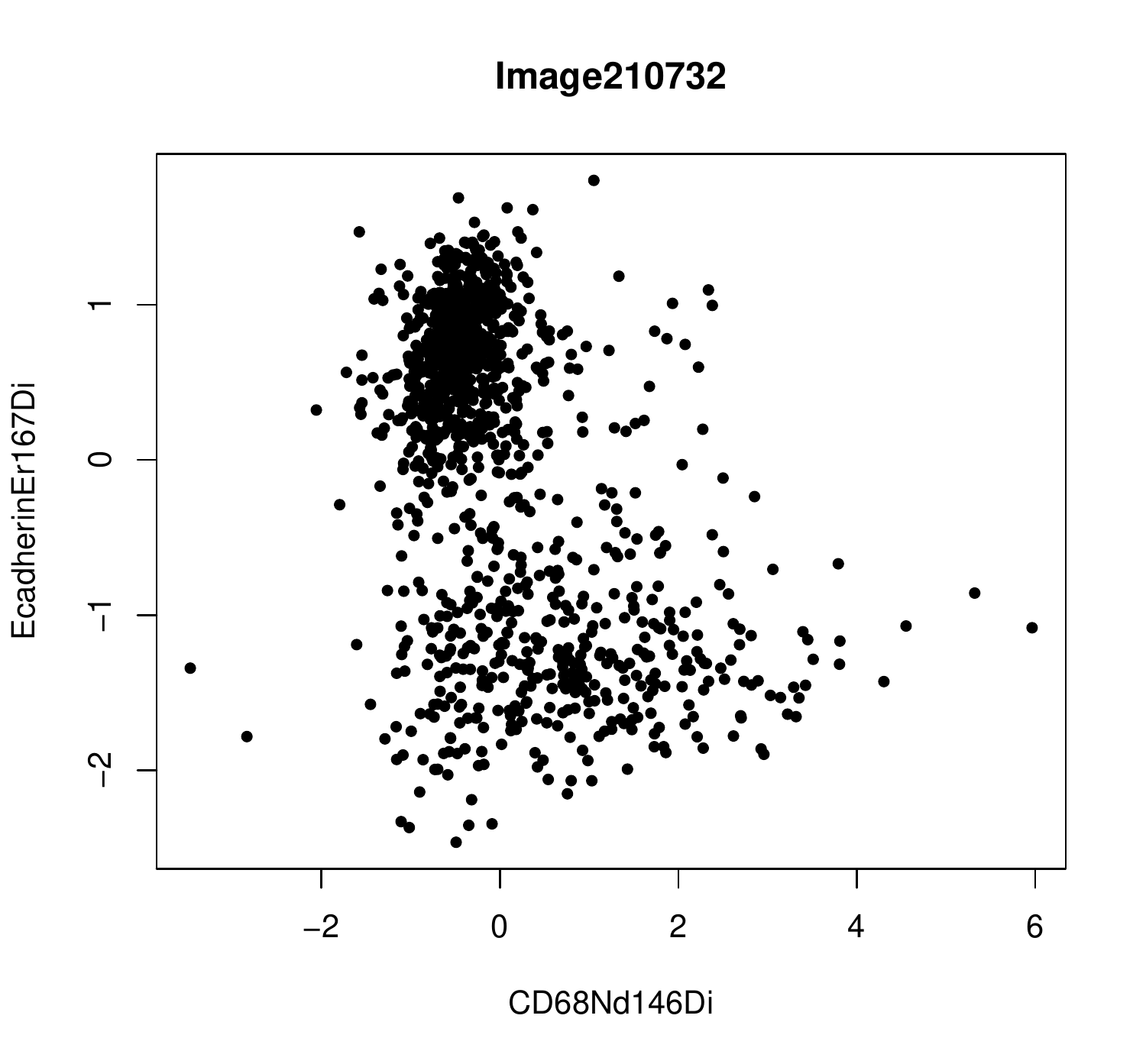}
	\includegraphics[angle=0,width=5cm]{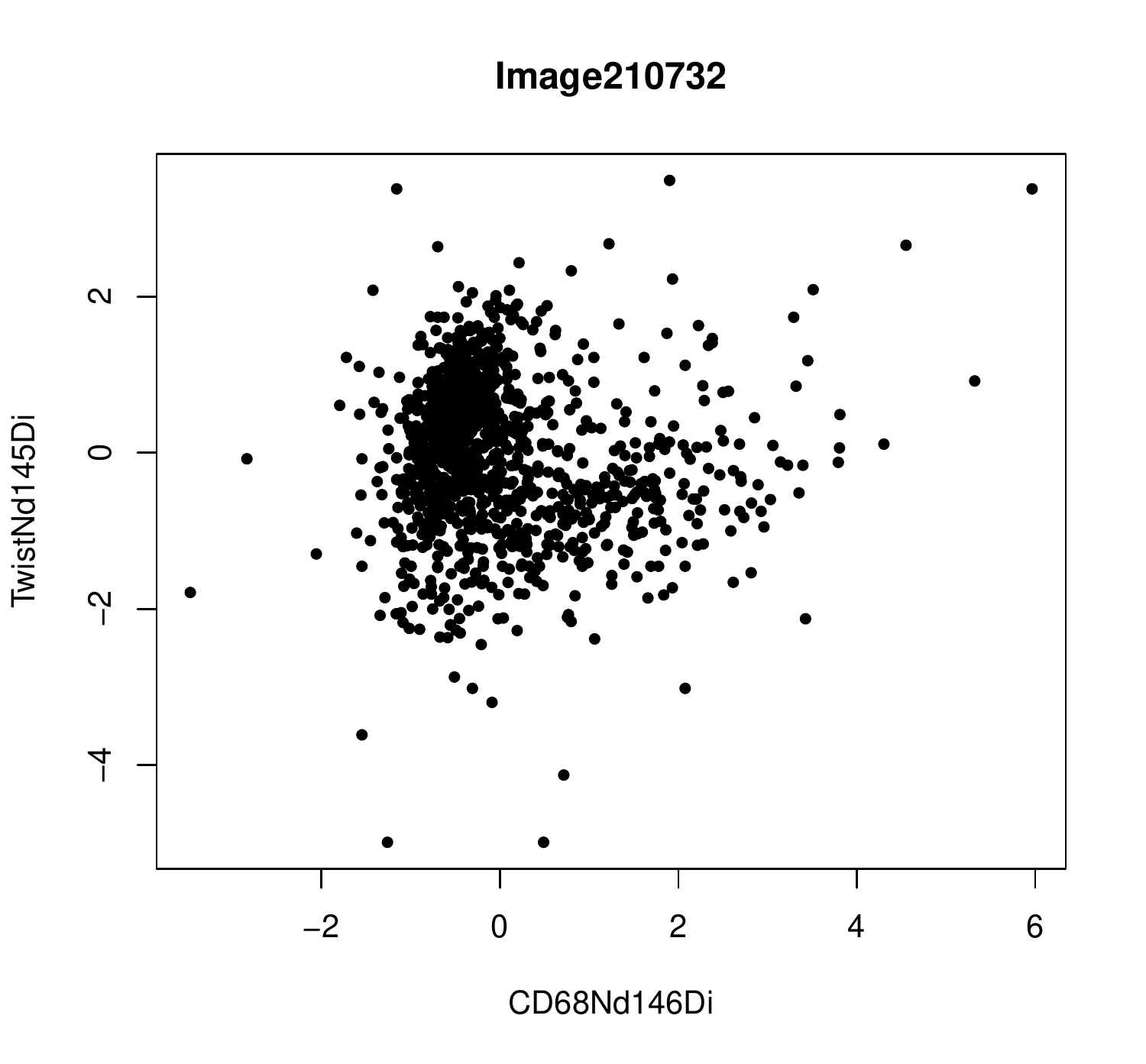}
	\includegraphics[angle=0,width=5cm]{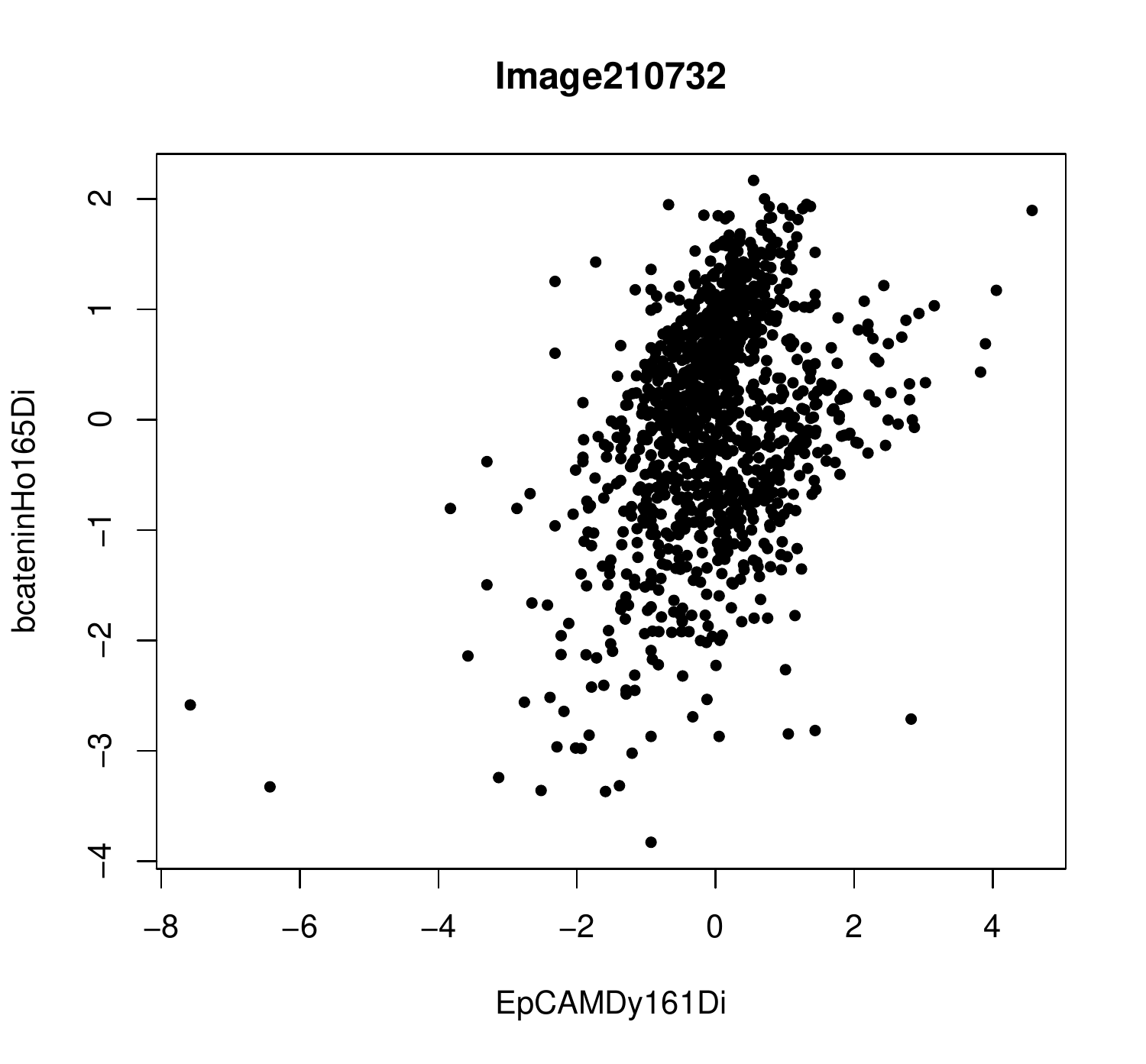}
	\includegraphics[angle=0,width=5cm]{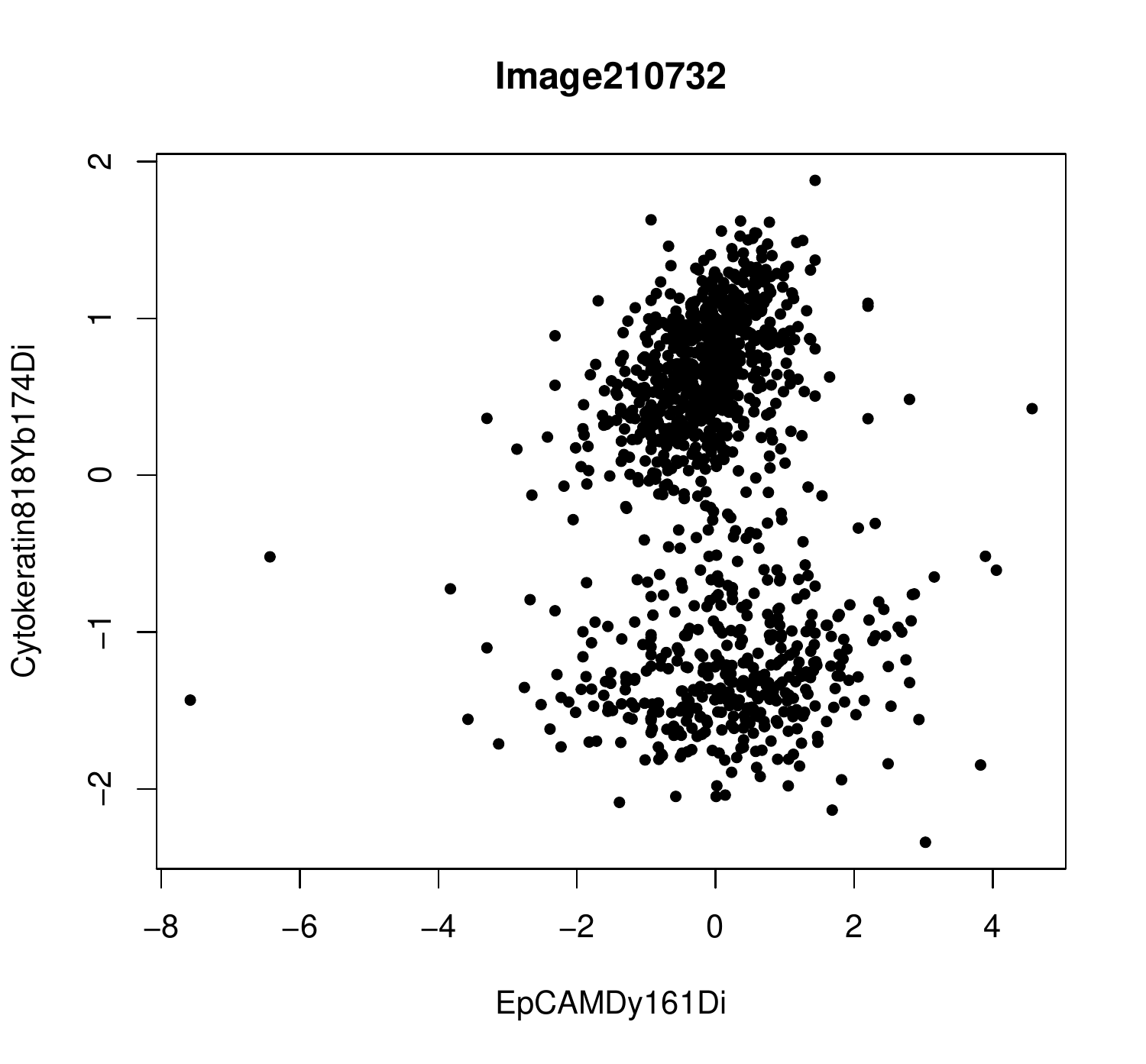}
	\includegraphics[angle=0,width=5cm]{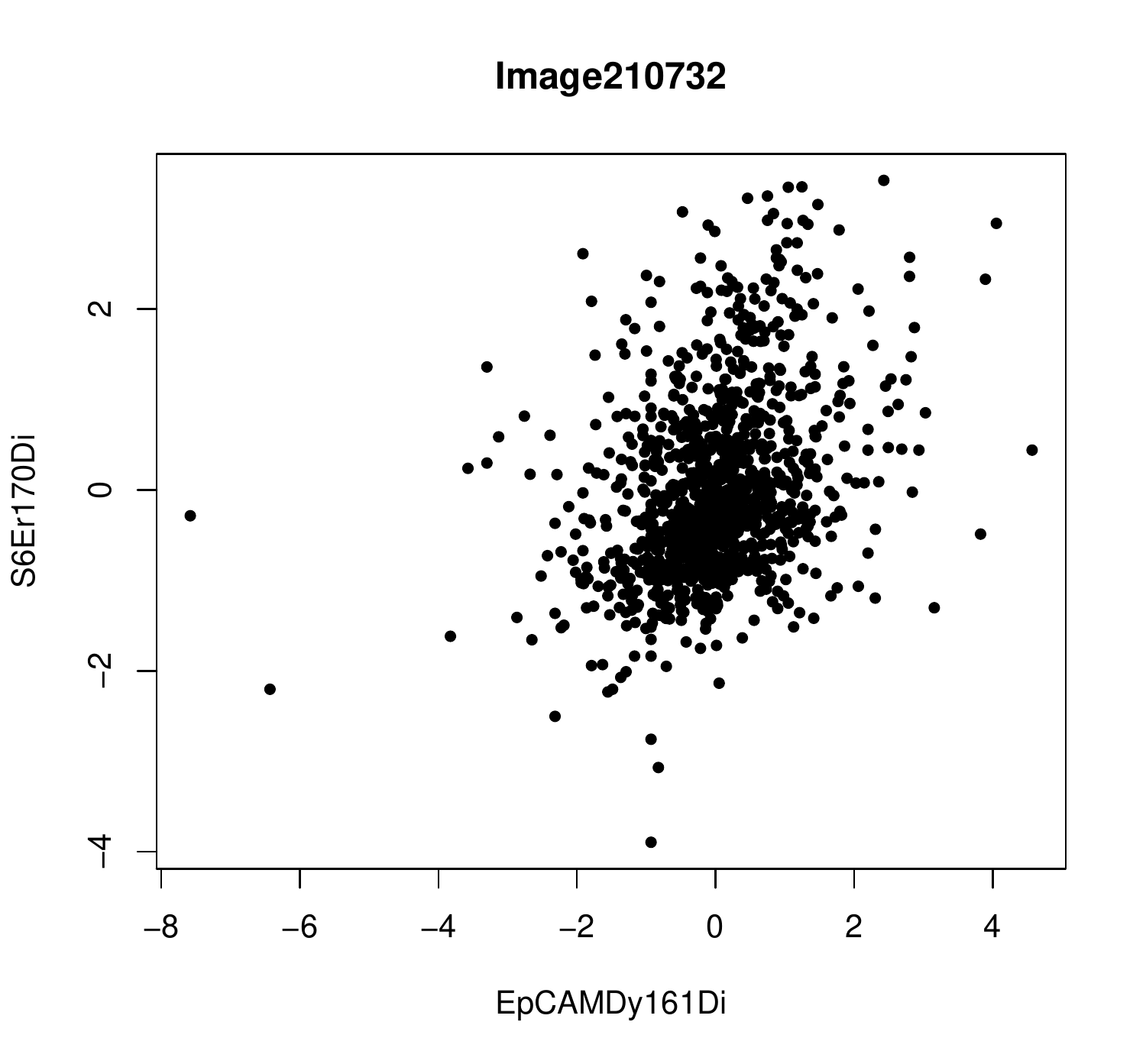}
	\includegraphics[angle=0,width=5cm]{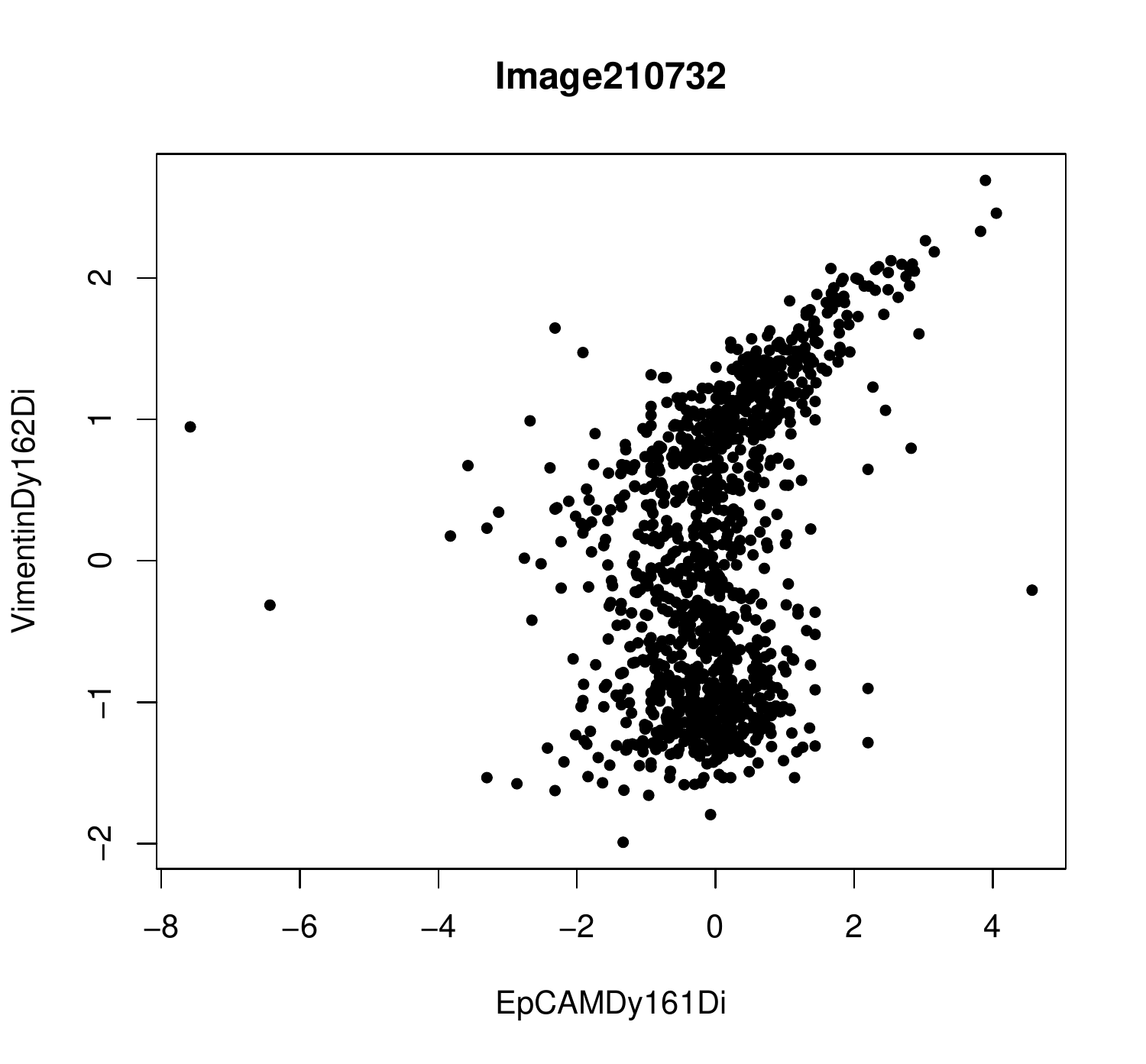}
	\includegraphics[angle=0,width=5cm]{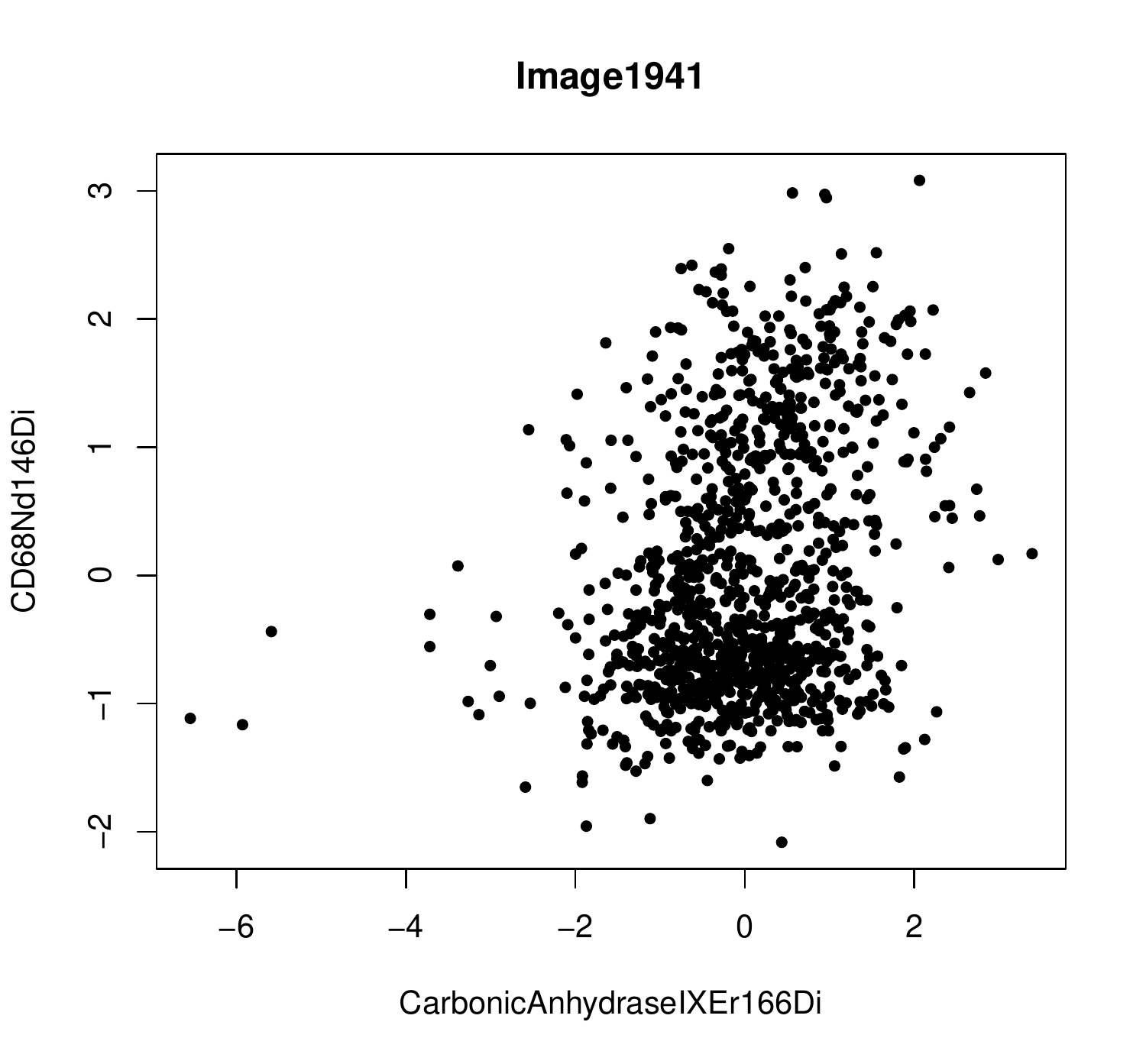}
	\includegraphics[angle=0,width=5cm]{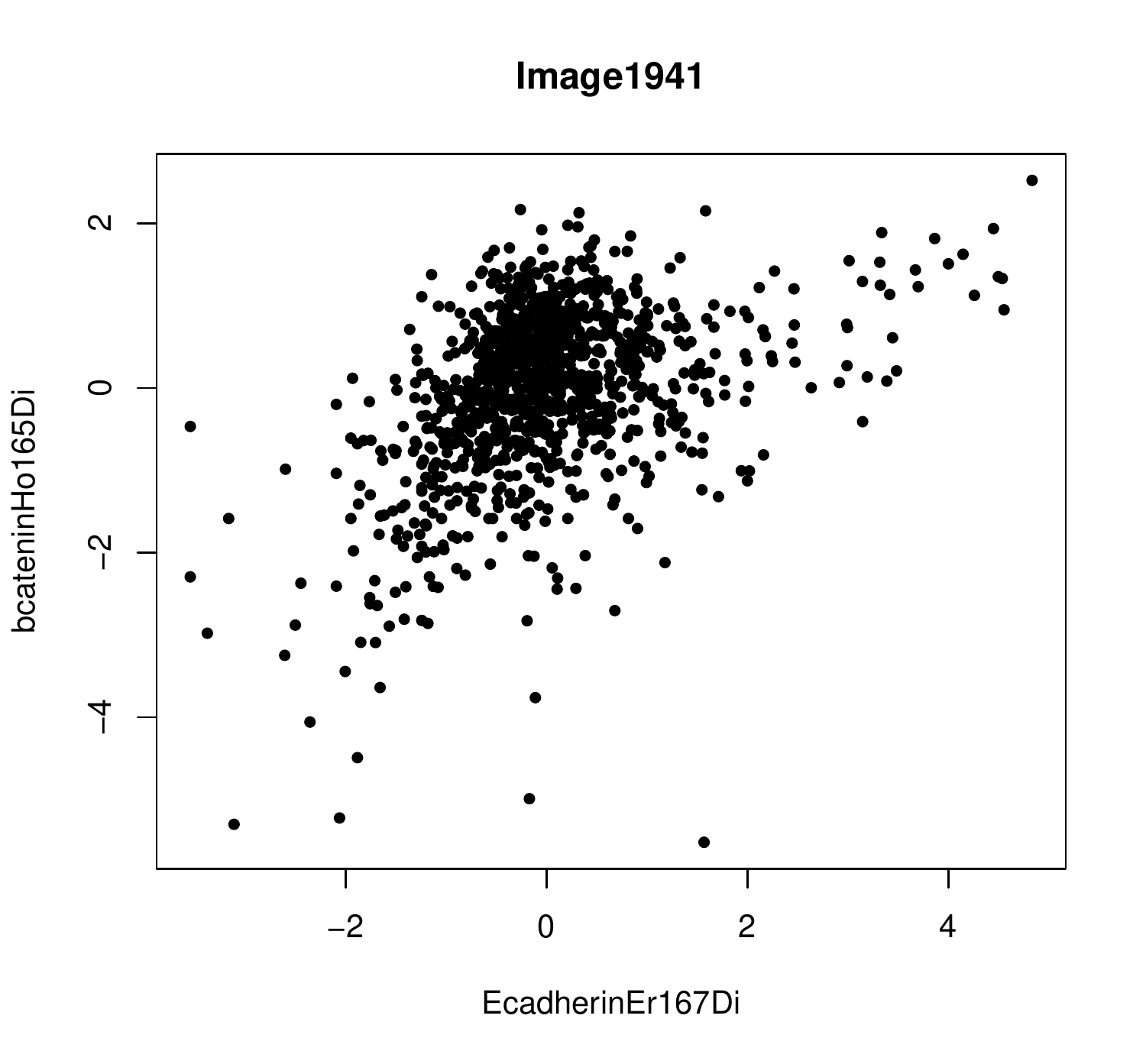}
	\includegraphics[angle=0,width=5cm]{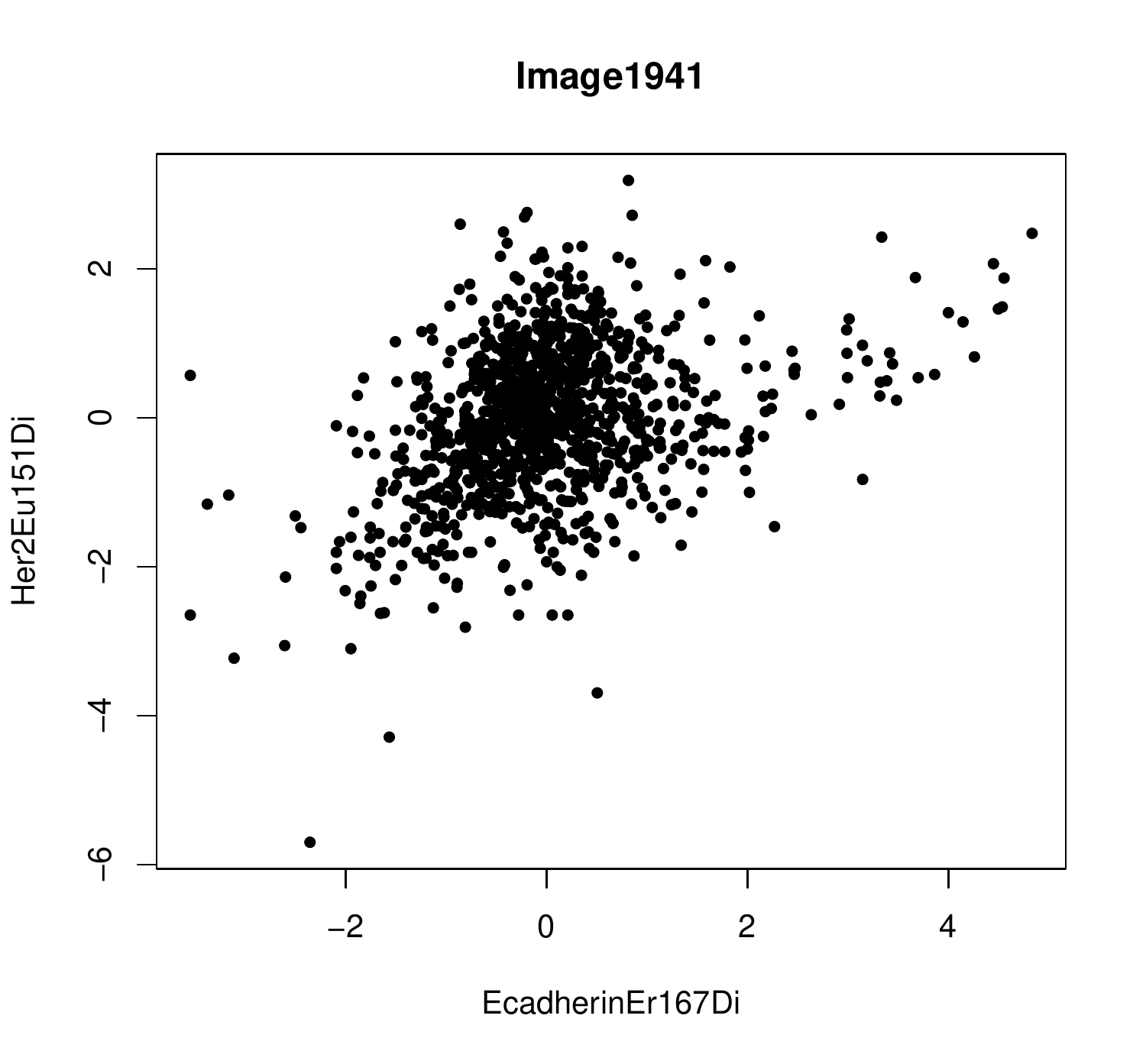}
	\includegraphics[angle=0,width=5cm]{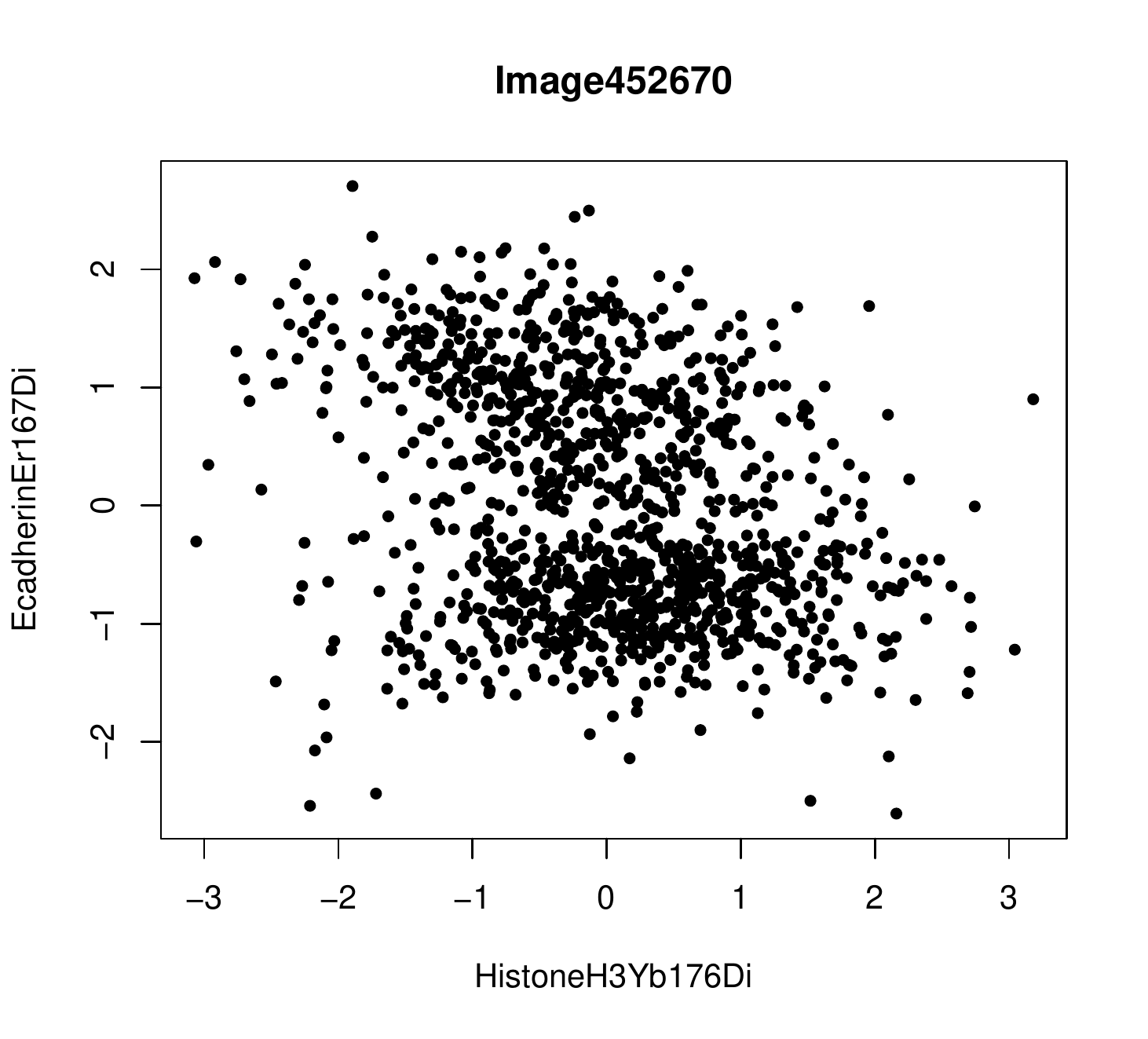}
	\includegraphics[angle=0,width=5cm]{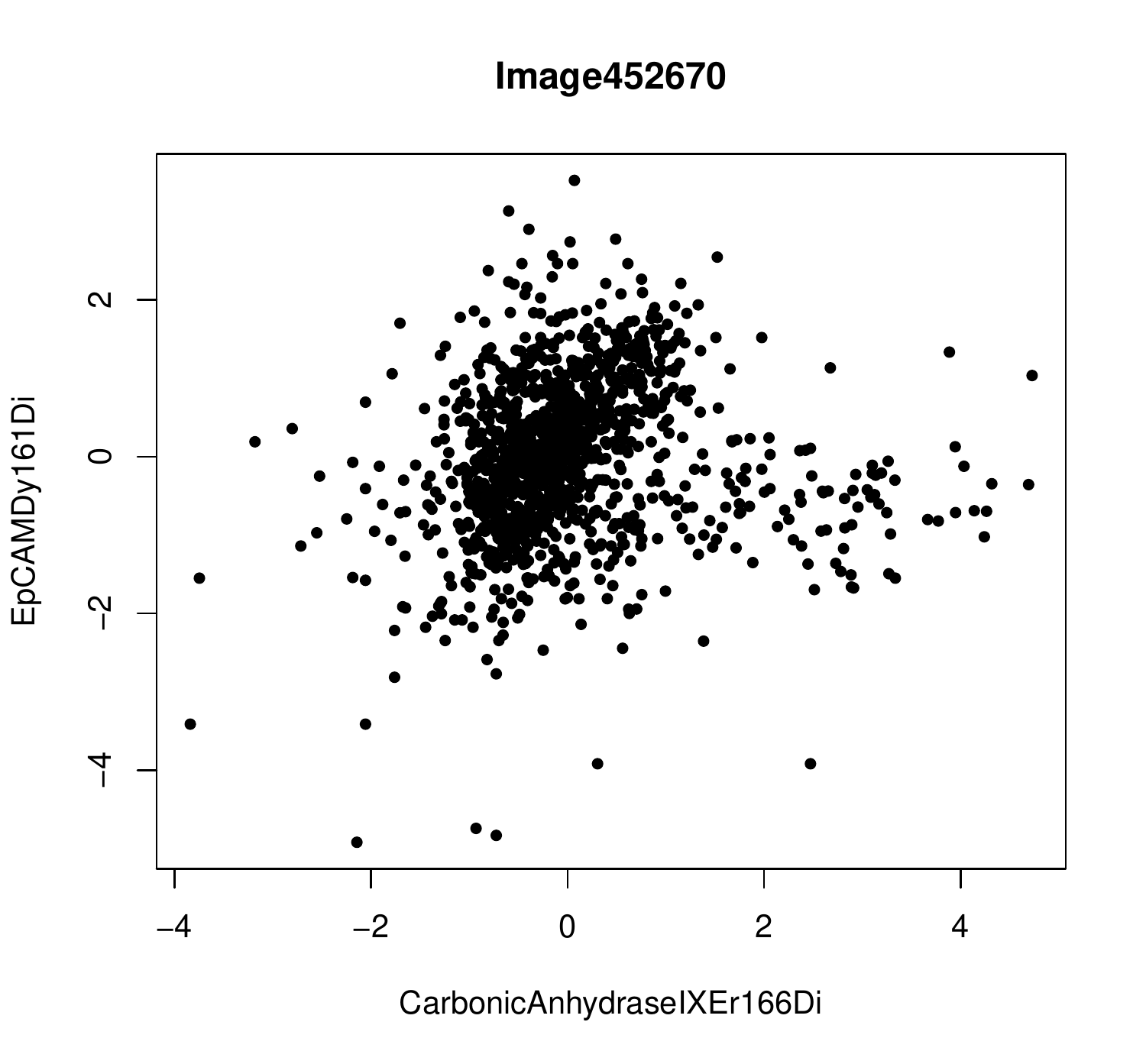}
	\includegraphics[angle=0,width=5cm]{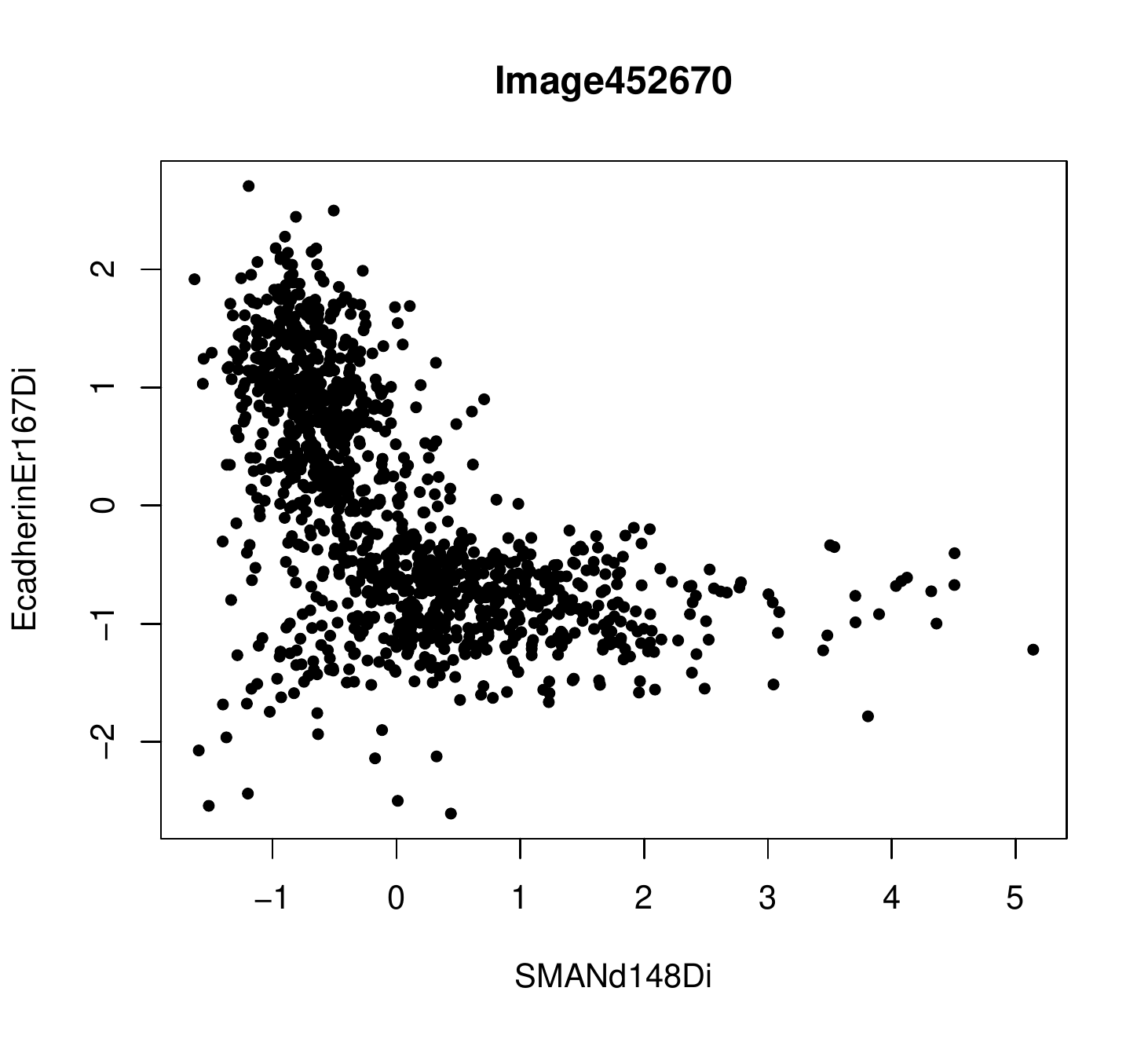}
	\caption{Pairwise scatter plots of epitope expressions for cells on different images, showing mixture of associations. } \label{scatter}
\end{figure}

\begin{figure}[h!]
	\centering
	\includegraphics[angle=0,width=7cm]{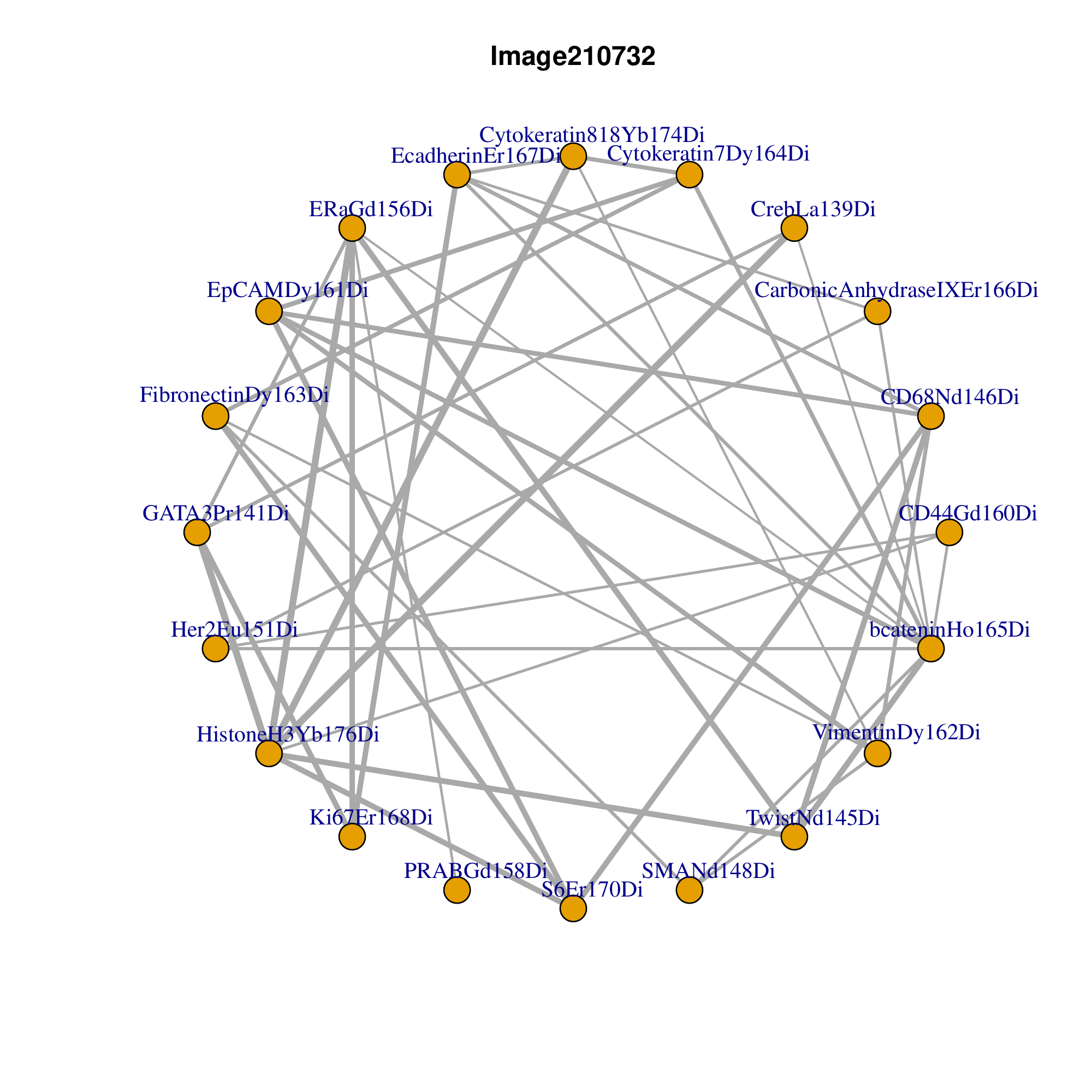}
	\includegraphics[angle=0,width=7cm]{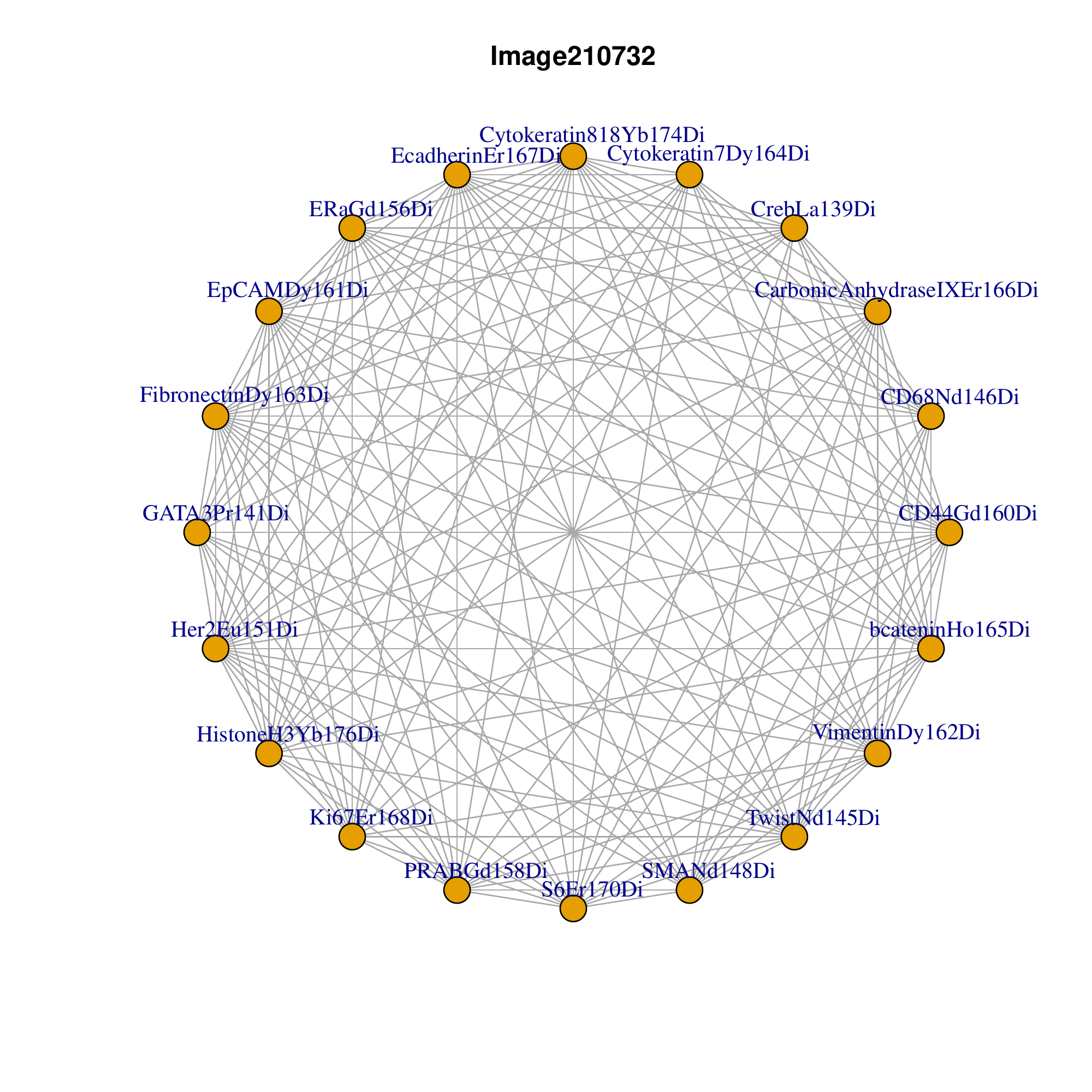}
	\includegraphics[angle=0,width=7cm]{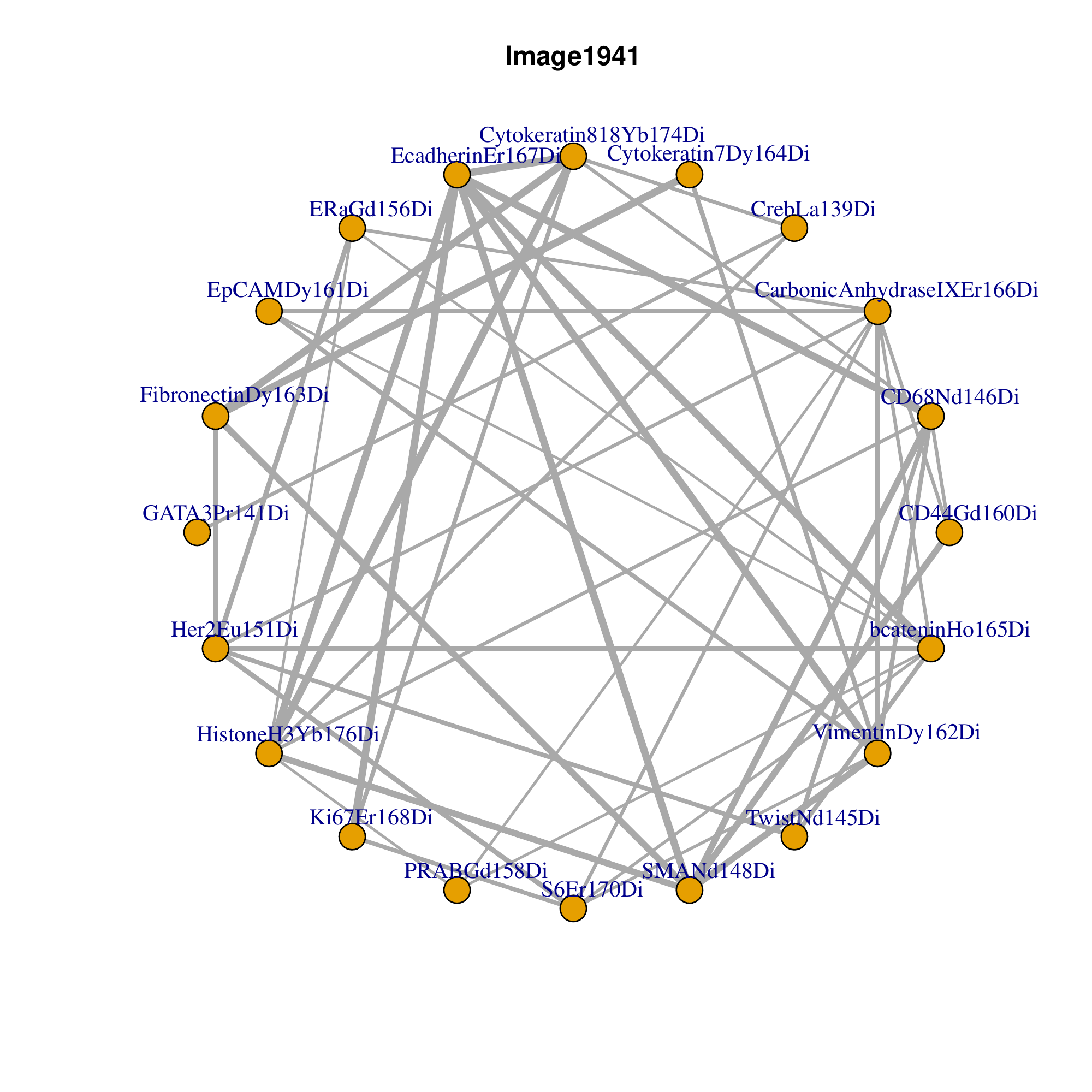}
	\includegraphics[angle=0,width=7cm]{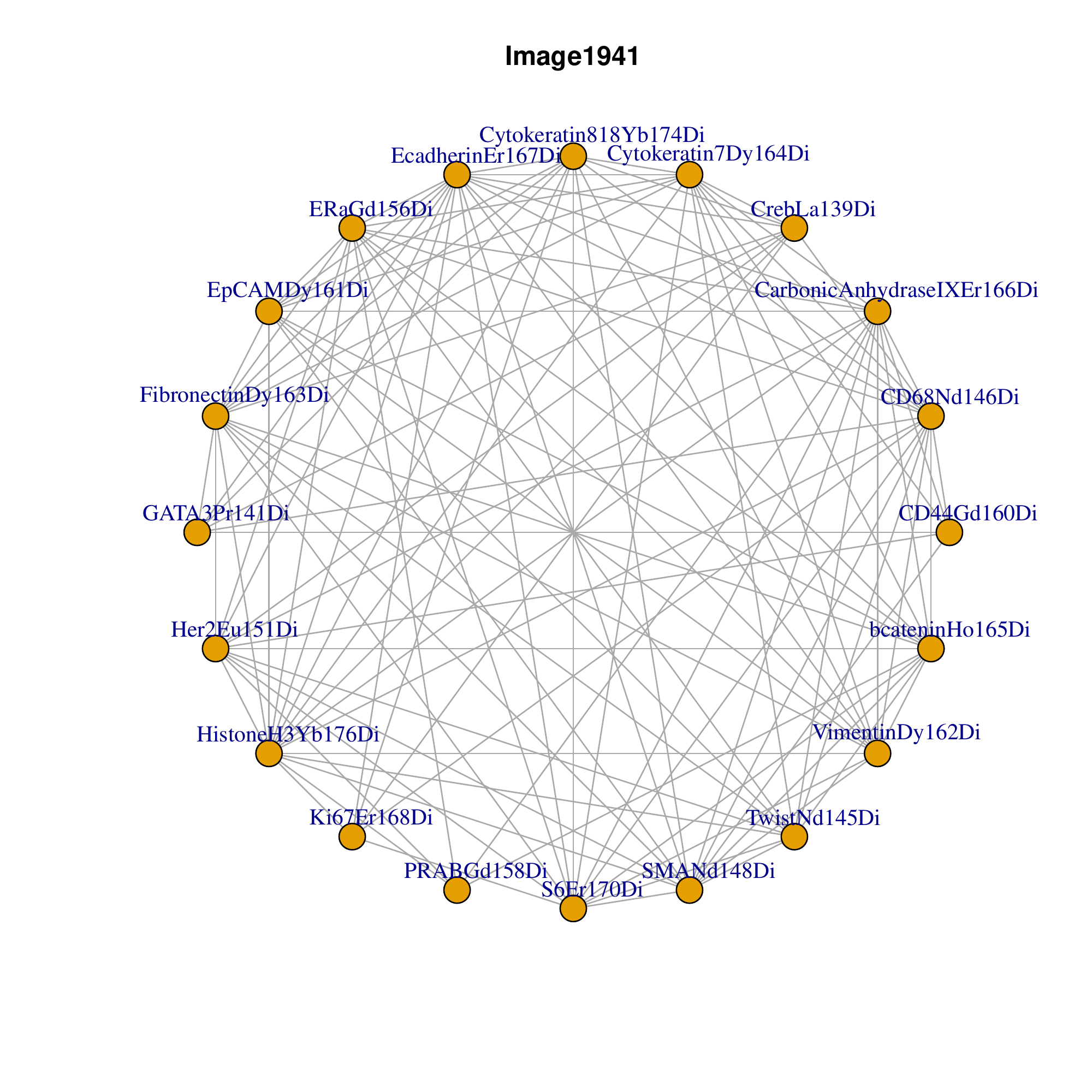}
	\includegraphics[angle=0,width=7cm]{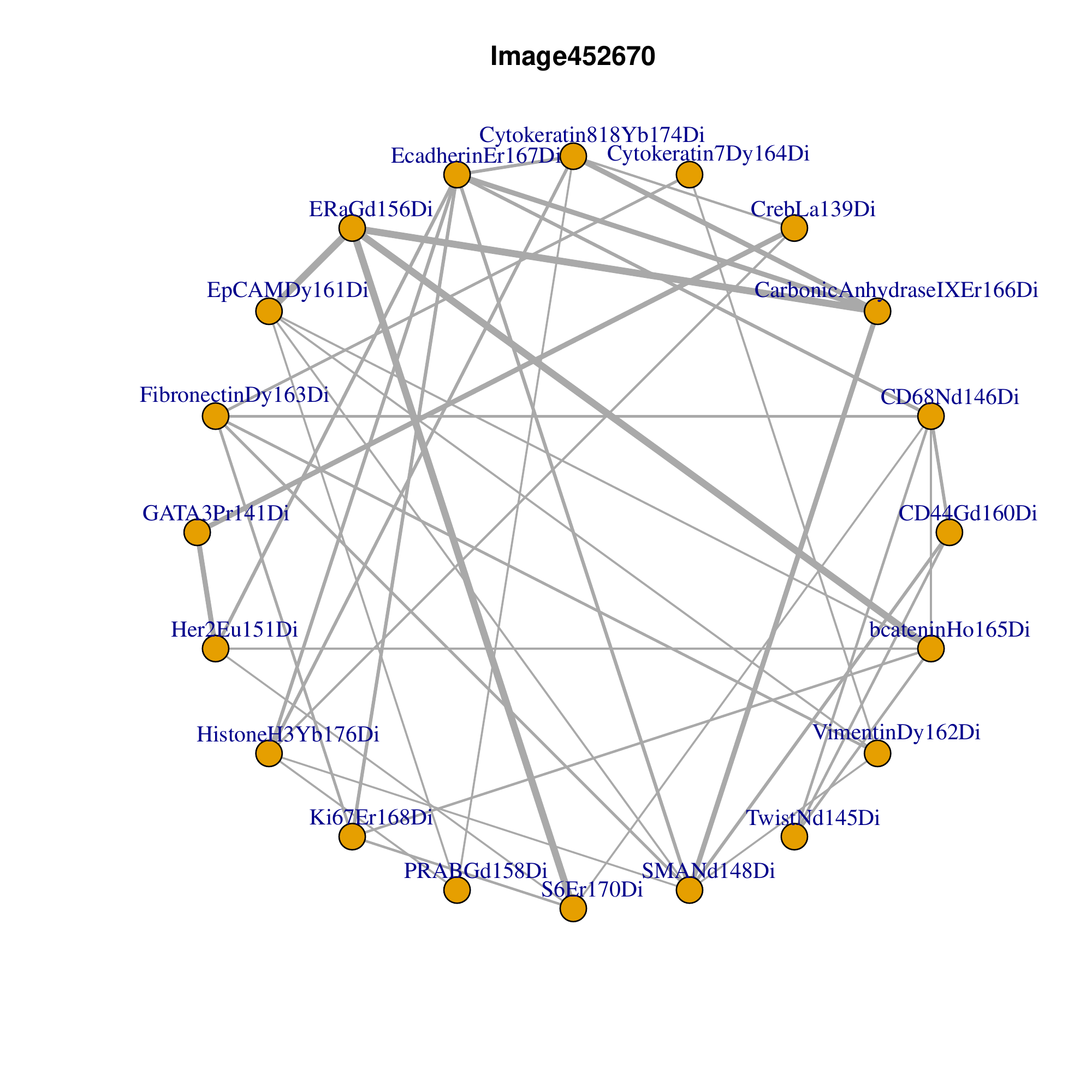}
	\includegraphics[angle=0,width=7cm]{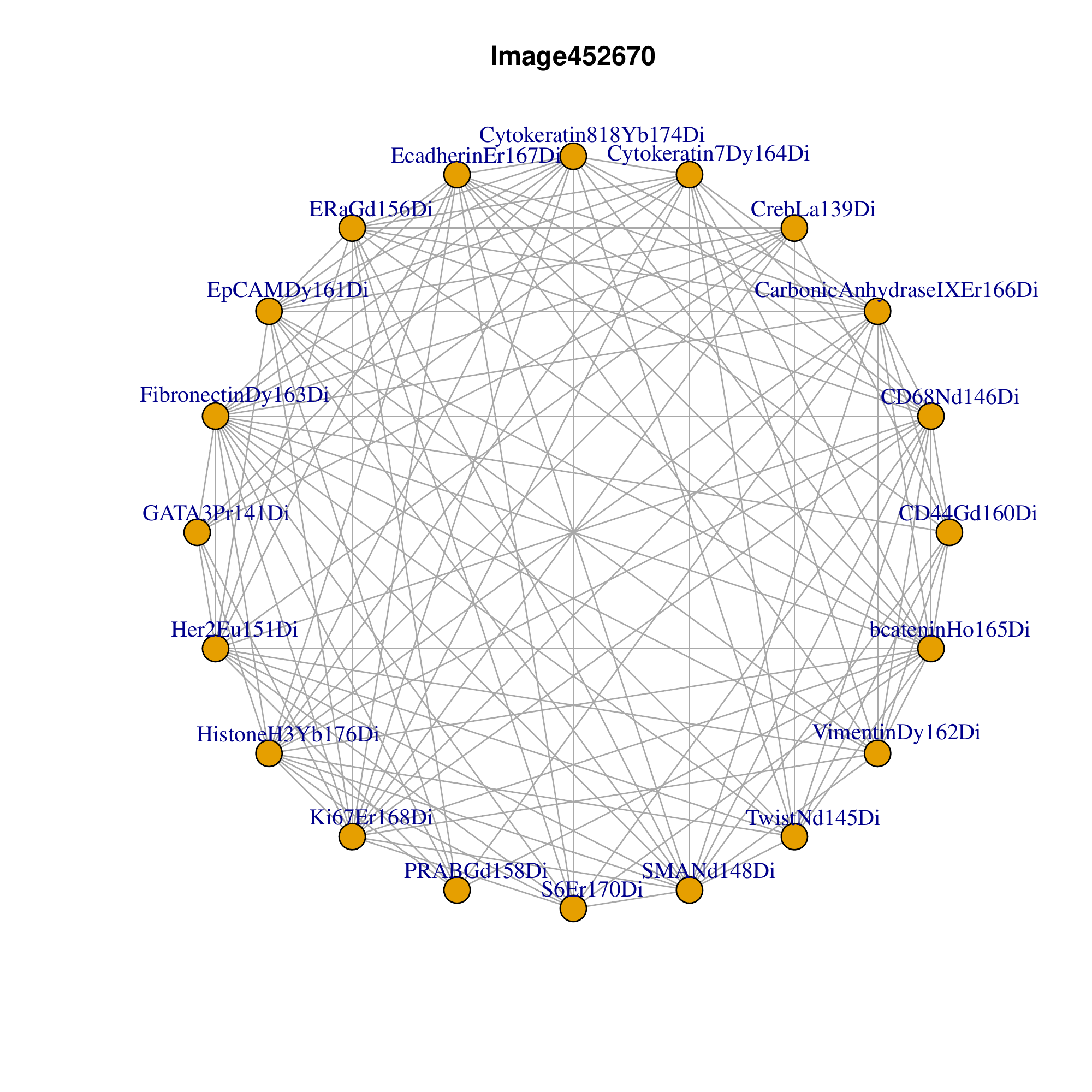}
	\caption{Networks of conditional dependence generated from node-wise regressions based on our proposed methods (left) and the standard Lasso based methods (right), both with FDR$<10\%$ for three different images. In the networks based on MLR, the widths of the edges were set to be proportional to the $\ell_2$ distances between the two mixed regression coefficients.} \label{network}
\end{figure}

As an example, we first focus on the Image 210732. In addition to the global heterogeneity, we also observed that the marginal associations between many pairs of epitopes or markers contain a two-class mixture pattern, as shown in Figure \ref{scatter}.
 To obtain a conditional dependence network based on the 1,151 cells in this image, we fitted node-wise mixed linear regressions and for each of them performed the proposed multiple testing procedure with FDR $<10\%$. The final network (Figure, \ref{network}, top left) was constructed such that the edges indicate the identified associations from at least one of such node-wise regressions. To better illustrate the effects of mixture, the widths of the edges were set to be proportional to the $\ell_2$ distances between the two mixed regression coefficients, so that a thicker edge indicates a larger discrepancy between the two mixtures.  As a comparison, we also obtained a network (Figure \ref{network}, top right) based on the standard node-wise Lasso and the multiple testing procedure of \cite{javanmard2019false} with the same FDR level. Similar to our simulation results, the standard Lasso-based methods tend to report many  more associations than our proposed method, due to its failure to account for the underlying mixtures and the resulting underestimated $p$-values for the individual tests. In particular, we found that many heterogeneous associations shown by scatter plots in Figure \ref{scatter} were indeed captured by our methods as the thicker edges in the network estimated by our mixture model.
 
Naturally, the above analysis can be conducted similarly for each of the images. Here we present the results for Image 1941 and Image 452670, as two additional examples. With the global heterogeneity shown in Figure \ref{tsne},  again we obtained  much denser networks from the standard Lasso based method and sparser networks from our proposed method (both with FDR $<10\%$ for the node-wise regressions). Moreover, many marginal associations (Figure \ref{scatter}) with  heterogeneous associations  had  thicker edges of the networks based on our proposed method. Our analysis also suggests that the naive application of Lasso to heterogeneous datasets can lead to false associations. 

\section{DISCUSSION}

 
 
The present paper introduced an iterative estimation procedure using an EM algorithm, a debiased approach for  individual coefficient inference based on the EM solutions, and a multiple testing procedure based on the debiased estimators for the high-dimensional mixed linear regression. 
Similar to many other works on EM algorithms \citep{balakrishnan2017statistical, yi2013alternating, wang2014high, yi2015regularized}, sample splitting  was used to facilitate the theoretical analysis in order to derive the estimation consistency.
However, the numerical results suggest that such data splitting seems to be unnecessary for achieving the desirable results in practice. It is interesting to develop novel technical tools for analyzing the algorithms without splitting the sample. 

The proposed EM algorithm assumes that the noise variance is known. Such an algorithm can be naturally extended to the case where the two noise variances are different and known. A more interesting problem is to develop an algorithm for the case where the two noise variances are different and unknown.
 In addition, this paper focuses on the two-class mixed regression model. It is interesting to extend the proposed algorithms to the general $k$-class mixed regression models and analyze its performance, especially when $k$ is unknown.

In addition to the estimation, individual coefficient inference and multiple testing problems considered in the current paper, there are several other interesting and related problems that are worth investigating. One such related problem is testing a single regression model against a  mixed regression model.  This involves, for example, the construction and analysis of a goodness-of-fit test. Finally, a natural generalization of the mixed linear regression model is the mixed generalized linear models (MGLM), where the outcome variables are allowed to be categorical. Estimation and multiple testing for high-dimensional MGLM are  important and challenging problems that we leave for future research. 

 \section{PROOFS}
We present in this section the proofs of Theorems \ref{EM:CI} and \ref{fdr}, the results on the individual coordinate inference and multiple testing. Theorem \ref{thm:inference} can be proved by using the same derivation as that in Theorem~\ref{EM:CI}, and the proofs Theorem~\ref{thm:estimation} and other technical lemmas are given in the Supplementary Materials \citep{supp}. 

\subsection{Proof of Theorem~\ref{EM:CI}}
We first state the following lemmas. 

\begin{lemma}\label{lemma:contraction2}
Under the same conditions as in Theorem~\ref{thm:estimation}. For any given vector $\bm m^*\in\R^p$, there exists a constant $C>0$ such that $$
\|\E[\frac{1}{n_T}\sum_{i=1}^{n_T}\gamma_{\bm\theta,i}^{(T)}\xx_i\xx_i^\top \bm m^*]-\E[\frac{1}{n_T}\sum_{i=1}^{n_T}\gamma_{\bm\theta^*,i}\xx_i\xx_i^\top \bm m^*]\|_2\le C\|\bm m^*\|_2 \|\bth^*-\hat\bth^{(T)}\|_2;
$$
$$
\|\E[\frac{1}{n_T}\sum_{i=1}^{n_T}\gamma_{\bm\theta,i}^{(T)}\xx_i\xx_i^\top \bm m^*]-\frac{1}{n_T}\sum_{i=1}^{n_T}\gamma_{\bm\theta,i}^{(T)}\xx_i\xx_i^\top \bm m^*\|_\infty=\|\bm m^*\|_2\cdot O_p(\sqrt\frac{\log p}{n_T}).
$$
\end{lemma}

%


\begin{lemma}\label{lemma:lbvariance2}
Under the same conditions as in Theorem~\ref{EM:CI}. There exists a constant $c>0$ such that $$
\bm m_j^\top (T_n(\hat\bth^{(T)}))_{1,1}\bm m_j\ge c, j\in[p].
$$
\end{lemma}

Given the lemmas, we now proceed to proving Theorem 2.  By symmetry, in the following, we only consider the case where $l=1$.

Recall that $n_T=n/T$ with $T\asymp \log n$, $\tilde\Sigma_{XX}=\frac{1}{n_T}\sum_{i=1}^{n_T}\xx_i\xx_i^\top$. We first verify that for $\mu=C\sqrt{\frac{\log p\log n}{n}}$ with sufficiently large constant $C$, the optimization of $\bm m_j$ is feasible, that is, there exits $\bm m_j^*\in\R^p$, such that $ ||\tilde\Sigma_{XX} \bm m_j^*- \bm e^{(p)}_j||_\infty\le \mu$ and $\|\bm m_j^*\|_1\le C\sqrt{\log n}$.

Take $\bm m_j^*=(\Sig^{-1})_j$ and use the fact that $\|\Sig^{-1}\|\le L$, we have some $\|\bm m_j^*\|_1\le C\sqrt{\log n}$.

Further, since $\|\bm m_j^*\|_2\le C\sqrt{\log n}$ and $n/T\asymp n/\log n$, by the Bernstein inequality and union bound, we get
$$
\|\E[\frac{1}{n_T}\sum_{i=1}^{n_T}\xx_i\xx_i^\top \bm m_j^*]-\frac{1}{n_T}\sum_{i=1}^{n_T}\xx_i\xx_i^\top \bm m_j^*\|_\infty=O_p(\sqrt\frac{\log p}{n}{\log n}).
$$

Therefore, the optimization is feasible, and recall that the solution is denoted as $\tilde{\bm m}_j$.
Then we proceed to showing the asymptotic normality. 

For a given $j$, by (3.2), we have
\begin{align}\label{decomposition2}
\sqrt{n_T}(\widehat{\bm \beta}_{1j}^u- \bbeta_{1j}^{*})= &\sqrt{n_T}(\bm m_j^\top\widehat\Sigma_{XY}-\bm m_j^\top\widehat\Sigma_{XX}{\bm \beta_1^*})+\sqrt{n_T}(\bm e_j^\top-\bm m_j^\top\widehat\Sigma_{XX})(\hat{\bm \beta}_1^{(T)}-{\bm \beta}_1^*)\\
\notag=&\sqrt{n_T}\bigg[\frac{1}{n_T}\sum_{i=1}^n\gamma_{\bm\theta,i}^{(T)}(y_i-\langle\xx_i,\bbeta_1^*\rangle)\bm m_j^\top\xx_i\bigg]+ \sqrt{n_T}\|\bm e_j^\top-\bm m_j^\top\widehat\Sigma_{XX} \|_\infty\cdot \|\hat{\bm \beta}_1^{(T)}-{\bm \beta}_1^* \|_1.
\end{align}

We then show that for $\bm m_j=\tilde{\bm m}_j/\hat\omega^{(T)}$, $\|\bm e_j^\top-\bm m_j^\top\widehat\Sigma_{XX} \|_\infty$ is small.

Recall that  $\widehat\Sigma_{XX}=\frac{1}{n_T}\sum_{i=1}^{n_T}\gamma_{\bm\theta,i}^{(T)}\xx_i\xx_i^\top$. By Lemma~\ref{lemma:contraction2}, we get $$
\|\E[\frac{1}{n_T}\sum_{i=1}^{n_T}\gamma_{\bm\theta,i}^{(T)}\xx_i\xx_i^\top \bm m_j]-\E[\frac{1}{n_T}\sum_{i=1}^{n_T}\gamma_{\bm\theta^*,i}\xx_i\xx_i^\top \bm m_j]\|_2\lesssim \|\bm m_j\|_2\cdot\|\bth^*-\hat\bth^{(T)}\|_2=O_p(\sqrt\frac{s\log p}{n}\log n),
$$
and
$$
\|\E[\frac{1}{n_T}\sum_{i=1}^{n_T}\gamma_{\bm\theta,i}^{(T)}\xx_i\xx_i^\top \bm m_j^*]-\frac{1}{n_T}\sum_{i=1}^{n_T}\gamma_{\bm\theta,i}^{(T)}\xx_i\xx_i^\top \bm m_j^*\|_\infty=O_p(\sqrt\frac{\log p}{n}\log n).
$$

Let $z_i$ be the class for the pair of data $(\xx_i, y_i)$, we obtain 
$$
\E[\frac{1}{n_T}\sum_{i=1}^{n_T}\gamma_{\bm\theta^*,i}\xx_i\xx_i^\top \bm m_j^*]=\E[\frac{1}{n_T}\sum_{i=1}^{n_T}\E_{\hat\bth^{*}}[1(z_i=1)\xx_i\xx_i^\top \bm m_j^*\mid \xx_i, y_i]]=\omega^* \bm\Sigma.
$$

Therefore, we have \begin{align*}
 &||\hat\Sigma_{XX} \bm m_j- \bm e^{(p)}_j||_\infty\\
 =& ||\E[\frac{1}{n_T}\sum_{i=1}^{n_T}\gamma_{\bm\theta,i}^{(T)}\xx_i\xx_i^\top \bm m_j] -\frac{1}{n_T}\sum_{i=1}^{n_T}\gamma_{\bm\theta,i}^{(T)}\xx_i\xx_i^\top \bm m_j||_\infty\\
&+ ||\E[\frac{1}{n_T}\sum_{i=1}^{n_T}\gamma_{\bm\theta,i}^{(T)}\xx_i\xx_i^\top \bm m_j]-\E[\frac{1}{n_T}\sum_{i=1}^{n_T}\gamma_{\bm\theta^*,i}\xx_i\xx_i^\top \bm m_j]||_\infty
+ ||\E[\frac{1}{n_T}\sum_{i=1}^n\gamma_{\bm\theta^*,i}\xx_i\xx_i^\top \bm m_j]- \bm e^{(p)}_j||_\infty\\
=&O_p(\sqrt\frac{\log p}{n}\log n)+O_p(\sqrt\frac{s\log p}{n}\log n)+\|\omega^*\Sig\bm m_j-\bm e_j^{(p)}\|_\infty\\
\le&O_p(\sqrt\frac{s\log p}{n}\log n)+\|\omega^*\Sig\bm m_j-\hat\omega^{(T)}\tilde\Sigma_{XX}\bm m_j\|_\infty+\|\hat\omega^{(T)}\tilde\Sigma_{XX}\bm m_j-\bm e_j^{(p)}\|_\infty\\
\le&O_p(\sqrt\frac{s\log p}{n}\log n)+\|\omega^*\Sig-\hat\omega^{(T)}\tilde\Sigma_{XX}\|_\infty\cdot\|\bm m_j\|_1+\|\tilde\Sigma_{XX}\tilde{\bm m}_j-\bm e_j^{(p)}\|_\infty\\
\le&O_p(\sqrt\frac{s\log p}{n}\log n).
\end{align*}

Then \eqref{decomposition2} becomes
\begin{align}
\notag&\sqrt{n_T}(\widehat{\bm \beta}_{1j}^u- \bbeta_{1j}^{*})\\
=&\sqrt{n_T}\bigg[\frac{1}{n_T}\sum_{i=1}^n\gamma_{\bm\theta,i}^{(T)}(y_i-\langle\xx_i,\bbeta_1^*\rangle)\bm m_j^\top\xx_i\bigg]+ \sqrt{n_T}\|\bm e_j^\top-\bm m_j^\top\widehat\Sigma_{XX} \|_\infty\cdot \|\hat{\bm \beta}_1^{(T)}-{\bm \beta}_1^* \|_1\\
\notag=&\langle \bm m_j, \frac{1}{\sqrt{n_T}}\sum_{i=1}^{n_T}\gamma_{\bm\theta,i}^{(T)}(y_i-\langle\xx_i,\bbeta_1^*\rangle)\xx_i\rangle+O_P(\frac{s^{3/2}\log p\log n}{\sqrt n})\\
\notag=&\langle \bm m_j, \frac{1}{\sqrt{n_T}}\sum_{i=1}^{n_T}\gamma_{\bm\theta,i}^{(T)}(y_i-\langle\xx_i,\bbeta_1^*\rangle)\xx_i\rangle+o_P(1).
\end{align}

Then, by Lemma~\ref{lemma:lbvariance2}, we have $$
\frac{\sqrt{n_T}\left(\widehat{\bm \beta}_{1 j}^u- \bbeta_{1 j}^{*}\right)}{\sqrt{\hat v_{1 j}}}=\frac{\langle \bm m_j, \frac{1}{\sqrt n}\sum_{i=1}^n\gamma_{\bm\theta,i}^{(T)}(y_i-\langle\xx_i,\bbeta_1^*\rangle)\xx_i\rangle}{\sqrt{\hat v_{1 j}}}+o_P(1).
$$
Finally,  using Lemma~\ref{lemma:score}, we obtain the desired result. $$
	\frac{\sqrt n\left(\widehat{\bm \beta}_{1 j}^u- \bbeta_{1 j}^{*}\right)}{\sqrt{\hat v_{1 j}}}\stackrel{d}{\to} N(0,1).
$$

\subsection{Proof of Theorem~\ref{fdr}}

We first consider the case when $\hat{t}$, given by \eqref{t.hat.fdr}, does not exist. In this case, we have $\hat{t}=\sqrt{2\log p}$. Note that for $j\in \HH_0$, we have
\[
T_j^{(1)}=\frac{\sqrt{n}\hat{\bbeta}_{1,j}^{u}}{\hat{v}^{(1)}_j}=\frac{\langle \bm m_j, \frac{1}{\sqrt n}\sum_{i=1}^n\gamma_{\bm\theta^*,i}(y_i-\langle\xx_i,\bbeta_{1}^*\rangle)\xx_i\rangle}{\hat{v}^{(1)}_j}+\frac{\sqrt{n}Rem_1}{\hat{v}^{(1)}_j},
\]
where $Rem_1=o_P(1/\sqrt{n})$, and a similar expression holds for $T_j^{(2)}$.
Then we have
\begin{align} \label{fdr.out}
&\Pro\bigg( \sum_{j\in\HH_0} I\big(|T_j|\ge \sqrt{2\log p} \big)\ge 1 \bigg)\nonumber\\
&\le \Pro\bigg( \sum_{j\in\HH_0} I\big(|T_j^{(1)}|\ge \sqrt{2\log p} \big)\ge 1 \bigg)+\Pro\bigg( \sum_{j\in\HH_0} I\big(|T_j^{(2)}|\ge \sqrt{2\log p} \big)\ge 1 \bigg)\nonumber \nonumber \\
&\le \Pro\bigg( \sum_{j\in\HH_0} I\bigg(\frac{\langle \bm m_j, \frac{1}{\sqrt n}\sum_{i=1}^n\gamma_{\bm\theta^*,i}(y_i-\langle\xx_i,\bbeta_{1}^*\rangle)\xx_i\rangle}{\hat{v}^{(1)}_j}+\frac{\sqrt{n}Rem_1}{\hat{v}^{(1)}_j}\ge \sqrt{2\log p} \bigg)\ge 1 \bigg)\nonumber\\
&\quad+\Pro\bigg( \sum_{j\in\HH_0} I\bigg(\frac{\langle \bm m_j, \frac{1}{\sqrt n}\sum_{i=1}^n\gamma_{\bm\theta^*,i}(y_i-\langle\xx_i,\bbeta_{1}^*\rangle)\xx_i\rangle}{\hat{v}^{(1)}_j}+\frac{\sqrt{n}Rem_1}{\hat{v}^{(1)}_j}\le -\sqrt{2\log p} \bigg)\ge 1 \bigg)\nonumber \\
&\quad +\Pro\bigg( \sum_{j\in\HH_0} I\bigg(\frac{\langle \bm m_j, \frac{1}{\sqrt n}\sum_{i=1}^n(1-\gamma_{\bm\theta^*,i})(y_i-\langle\xx_i,\bbeta_{2}^*\rangle)\xx_i\rangle}{\hat{v}^{(2)}_j}+\frac{\sqrt{n}Rem_2}{\hat{v}^{(2)}_j}\ge \sqrt{2\log p} \bigg)\ge 1 \bigg)\nonumber\\
&\quad+\Pro\bigg( \sum_{j\in\HH_0} I\bigg(\frac{\langle \bm m_j, \frac{1}{\sqrt n}\sum_{i=1}^n(1-\gamma_{\bm\theta^*,i})(y_i-\langle\xx_i,\bbeta_{2}^*\rangle)\xx_i\rangle}{\hat{v}^{(2)}_j}+\frac{\sqrt{n}Rem_2}{\hat{v}^{(2)}_j}\le -\sqrt{2\log p} \bigg)\ge 1 \bigg).
\end{align}
Define $(v^{(1)}_j)^2=\Var(\langle \bm m_j, \frac{1}{\sqrt n}\sum_{i=1}^n\gamma_{\bm\theta^*,i}(y_i-\langle\xx_i,\bbeta_{1}^*\rangle)\xx_i\rangle )$. For any $\epsilon>0$, we can bound the first term by
\begin{align*}
&\Pro\bigg( \sum_{j\in\HH_0} I\bigg(\frac{\langle \bm m_j, \frac{1}{\sqrt n}\sum_{i=1}^n\gamma_{\bm\theta^*,i}(y_i-\langle\xx_i,\bbeta_{1}^*\rangle)\xx_i\rangle}{\hat{v}^{(1)}_j}+\frac{\sqrt{n}Rem_1}{\hat{v}^{(1)}_j}\ge \sqrt{2\log p} \bigg)\ge 1 \bigg)\\
&=\Pro\bigg( \sum_{j\in\HH_0} I\bigg(\frac{\langle \bm m_j, \frac{1}{\sqrt n}\sum_{i=1}^n\gamma_{\bm\theta^*,i}(y_i-\langle\xx_i,\bbeta_{1}^*\rangle)\xx_i\rangle}{{v}^{(1)}_j}\ge \frac{\hat{v}^{(1)}_j}{v^{(1)}_j}\sqrt{2\log p}-\frac{\sqrt{n}Rem_1}{{v}^{(1)}_j} \bigg)\ge 1 \bigg)\\
&\le \Pro\bigg( \sum_{j\in\HH_0} I\bigg(\frac{\langle \bm m_j, \frac{1}{\sqrt n}\sum_{i=1}^n\gamma_{\bm\theta^*,i}(y_i-\langle\xx_i,\bbeta_{1}^*\rangle)\xx_i\rangle}{{v}^{(1)}_j}\ge (1-\epsilon)\sqrt{2\log p}-\epsilon\bigg)\ge 1\bigg)\\
&\quad +\Pro\bigg( \max_{j\in \HH_0} \bigg|\frac{\sqrt{n}Rem_1}{{v}^{(1)}_j} \bigg|\ge\epsilon \bigg)+\Pro\bigg( \bigg|\frac{\hat{v}^{(1)}_j}{v^{(1)}_j}-1  \bigg|\ge \epsilon \bigg)\\
&\le p\max_{j\in\HH_0} \Pro\bigg( \frac{\langle \bm m_j, \frac{1}{\sqrt n}\sum_{i=1}^n\gamma_{\bm\theta^*,i}(y_i-\langle\xx_i,\bbeta_{1}^*\rangle)\xx_i\rangle}{{v}^{(1)}_j}\ge (1-\epsilon)\sqrt{2\log p}-\epsilon\bigg)+\Pro\bigg( \max_{j\in \HH_0} \bigg|\frac{\sqrt{n}Rem_1}{{v}^{(1)}_j} \bigg|\ge\epsilon \bigg)\\
&\quad +\Pro\bigg( \bigg|\frac{\hat{v}^{(1)}_j}{v^{(1)}_j}-1  \bigg|\ge \epsilon \bigg).
\end{align*}
By the proof of Theorem~\ref{EM:CI}, we know that
\[
\Pro\bigg( \max_{j\in \HH_0}\bigg|\frac{\sqrt{n}Rem_1}{v^{(1)}_j}\bigg|\ge\epsilon \bigg)\to 0,\qquad  \Pro\bigg( \bigg|\frac{\hat{v}^{(1)}_j}{v^{(1)}_j}-1  \bigg|\ge \epsilon \bigg)\to 0.
\]
In addition, for $j\in \HH_0$, let 
\[
T_{0j}^{(1)}=\frac{\langle \bm m_j, \frac{1}{\sqrt n}\sum_{i=1}^n\gamma_{\bm\theta^*,i}(y_i-\langle\xx_i,\bbeta_{1}^*\rangle)\xx_i\rangle}{{v}^{(1)}_j}.
\]
where {$\E\langle \bm m_j,\gamma_{\bm\theta^*,i}(y_i-\langle\xx_i,\bbeta_{1}^*\rangle)\xx_i\rangle/v^{(1)}_j=0$ and $\text{Var}(\E\langle \bm m_j,\gamma_{\bm\theta^*,i}(y_i-\langle\xx_i,\bbeta_{1}^*\rangle)\xx_i\rangle/v^{(1)}_j)=1$}. Conditional on $X$ by Lemma 6.1 of \cite{liu2013gaussian}, we have
\beq \label{5.13}
\sup_{0\le t\le 4\sqrt{\log p}}\bigg| \frac{P(|T_{0j}^{(1)}|\ge t)}{G(t)}-1\bigg|\le C(\log p)^{-1}.
\eeq
Hereafter, unless explicitly noted, all of our discussion will be conditional on $\{\xx_i\}_{i=1}^n$.
Now let $t=(1-\epsilon)\sqrt{2\log p}-2\epsilon$, we have
\[
\Pro\bigg( T_{0j}^{(1)}\ge (1-\epsilon)\sqrt{2\log p}-2\epsilon\bigg)\le G((1-\epsilon)\sqrt{2\log p}-2\epsilon)+C\frac{G((1-\epsilon)\sqrt{2\log p}-2\epsilon)}{\log p}.
\]
Hence
\[
p\max_{j\in\HH_0} \Pro\bigg( T_{0j}\ge (1-\epsilon)\sqrt{2\log p}-\epsilon\bigg)\le CpG((1-\epsilon)\sqrt{2\log p}-2\epsilon)+O(p^{-c}),
\]
which goes to zero as $(n,p)\to\infty$. By symmetry, we know that the rest three terms in (\ref{fdr.out}) also goes to 0. Therefore we have proved the theorem when $\hat{t}=\sqrt{2\log p}$.

Now consider the case when $0\le \hat{t}\le b_p$ holds. We have
\begin{align*}
\text{FDP}(\hat{t})&= \frac{\sum_{j\in \mathcal{H}_0} I\{ |T_{j}| \ge \hat{t}\}}{\max \big\{\sum_{j=1}^p I\{ |T_{j}| \ge \hat{t}\},1   \big\}}\le \frac{\sum_{j\in \mathcal{H}_0} I\{ |T_{j}^{(1)}| \ge \hat{t}\}+\sum_{j\in \mathcal{H}_0} I\{ |T_{j}^{(2)}| \ge \hat{t}\}}{\max \big\{\sum_{j=1}^p I\{ |T_{j}| \ge \hat{t}\},1   \big\}}.
\end{align*}
Note that for $\ell=1,2$,
\[
\frac{\sum_{j\in \mathcal{H}_0} I\{ |T_{j}^{(\ell)}| \ge \hat{t}\}}{\max \big\{\sum_{j=1}^p I\{ |T_{j}| \ge \hat{t}\},1   \big\}}\le \frac{p_0G(\hat{t}) }{\max \big\{\sum_{j=1}^p I\{ |T_{j}| \ge \hat{t}\},1   \big\}}(1+A^{(\ell)}_p)
\]
where
\[
A_p^{(\ell)}=\sup_{0\le t\le b_p}\bigg|\frac{\sum_{j\in \mathcal{H}_0} I\{ |T_{j}^{(\ell)}| \ge t\}}{p_0G(t)} -1\bigg|.
\]
Note that by definition
\[
\frac{p_0G(\hat{t}) }{\max \big\{\sum_{j=1}^p I\{ |T_{j}| \ge \hat{t}\},1   \big\}}\le \frac{p_0\alpha}{p}.
\]
The proof is complete if $A_p^{(\ell)}\to 0$ in probability. The rest of the proof is devoted to it. We first show that
\beq \label{TT0}
|T_j^{(\ell)}-T_{0j}^{(\ell)}|=o_P(1/\sqrt{\log p}).
\eeq
To see this, we notice that, under the sparsity condition { $s=o\big(\frac{ n^{1/2}}{\log^{3/2}p \log n}\big)$}, with probability at least $1-O(p^{-c})$,
\begin{align*}
|T_j^{(\ell)}-T_{0j}^{(\ell)}|&\le \bigg|\frac{\sqrt{n}Rem_\ell}{{v}^{(\ell)}_j}\bigg|\cdot\bigg|\frac{v_j^{(\ell)}}{\hat{v}^{(\ell)}_j} \bigg|+\bigg|T^{(\ell)}_{0j}({v}^{(\ell)}_j/\hat{v}^{(\ell)}_j-1)\bigg|= o\bigg( \frac{1}{\sqrt{\log p}}\bigg).
\end{align*}
By the fact that $G(t+o(1/\sqrt{\log p}))/G(t) = 1+o(1)$ uniformly in $0\le t\le \sqrt{2\log p}$, it suffices to show that
\beq \label{5.14}
\sup_{0\le t\le b_{p}} \bigg|  \frac{\sum_{j\in \HH_0} I\{|T_{0j}^{(\ell)}| \ge t \}}{p_0G(t)}-1 \bigg|\rightarrow 0\quad \text{ in probability.}
\eeq
Let $z_0<z_1<...<z_{d_p}\le 1$ and $t_i = G^{-1}(z_i)$, where $z_0 = G(b_p)$, $z_i = c_p/p+c_p^{2/3}e^{i^{\delta}}/p$ with $c_p=pG(b_p)$, and $d_p = [\log ((p-c_p)/c_p^{2/3})]^{1/\delta}$ and $0<\delta<1$, which will be specified later. We have $G(t_i)/G(t_{i+1}) = 1+o(1)$ uniformly in $i$, and $t_0/\sqrt{2\log(p/c_p)}=1+o(1)$. Note that uniformly for $1\le j\le m$, $G(t_i)/G(t_{i-1})\to 1$ as $p\to \infty$. The proof of (\ref{5.14}) reduces to show that
\beq \label{5.15}
\max_{0\le i\le d_{p}} \bigg|  \frac{\sum_{j\in \HH_0} I\{|T^{(\ell)}_{0j}| \ge t_i \}}{p_0G(t_i)}-1 \bigg|\rightarrow 0
\eeq
in probability. Hereafter, we omit the dependence on the index $\ell$ for simplicity. In fact, for each $\epsilon >0$, we have
\begin{align*}
&\Pro\bigg(   \max_{0\le i\le d_{p}} \bigg| \frac{\sum_{j\in \HH_0} [I\{|T_{0j}| \ge t_i \}-G(t_i) ]}{p_0 G(t_i)} \bigg|\ge \epsilon \bigg)  \le \sum_{j=0}^{d_p} \Pro\bigg(   \bigg| \frac{\sum_{j\in \HH_0} [I\{|T_{0j}| \ge t_i \}-G(t_i) ]}{p_0 G(t_i)} \bigg|\ge \epsilon/2 \bigg).
\end{align*}
Set $I(t) = \frac{\sum_{j\in \HH_0} [I\{|T_{0j}| \ge t \}-P(|T_{0j}| \ge t)]}{p_0 G(t)}.$
By Markov's inequality $P(|I(t_i) |\ge \epsilon /2) \le \frac{\E [I(t_i)]^2 }{\epsilon^2/4},$
and it suffices to show $\sum_{j=0}^{d_p} \E [I(t_i)]^2=o(1)$.
To see this, by (\ref{5.13}),
\begin{align*}
\E I^2(t) &= \frac{\sum_{j\in \HH_0} [P(|T_{0j}| \ge t)-P^2(|T_{0j}| \ge t)]}{p_0^2G^2(t)}\\
&\quad + \frac{\sum_{j,k\in \HH_0, k\ne j} [P(|T_{0k}| \ge t,|T_{0j}| \ge t)-P(|T_{0k}| \ge t)P(|T_{0j}| \ge t)]}{p_0^2G^2(t)}\\
& \le \frac{C}{p_0G(t)} + \frac{1}{p_0^2}\sum_{(j,k)\in\mathcal{A}(\epsilon)\cap\HH_0} \frac{P(|T_{0k}| \ge t,|T_{0j}| \ge t)}{G^2(t)}\\
&\quad+  \frac{1}{p_0^2}\sum_{(j,k)\in\mathcal{A}(\epsilon)^c \cap \HH_0} \bigg[\frac{P(|T_{0k}| \ge t,|T_{0j}| \ge t)}{G^2(t)} -1\bigg]\\
&= \frac{C}{p_0G(t)} + I_{11}(t) + I_{12}(t).
\end{align*}
For $(j,k)\in\mathcal{A}(\epsilon)^c \cap \HH_0$, applying Lemma 6.1 in \cite{liu2013gaussian}, we have $I_{12}(t) \le C(\log p)^{-1-\xi}$
for some $\xi>0$ uniformly in $0<t<\sqrt{2\log p}$. By Lemma 6.2 in \cite{liu2013gaussian}, for $(j,k)\in\mathcal{A}(\epsilon)\cap \HH_0$, we have
\[
P(|T_{0k}| \ge t,|T_{0j}| \ge t) \le C(t+1)^{-2}\exp \bigg(-\frac{t^2}{1+|\rho_{jk}|} \bigg).
\]
So that
\[
I_{11}(t) \le C \frac{1}{p_0^2}\sum_{(j,k)\in\mathcal{A}(\epsilon)\cap\HH_0}(t+1)^{-2}\exp \bigg(-\frac{t^2}{1+|\rho_{jk}|} \bigg)G^{-2}(t) \le C \frac{1}{p_0^2}\sum_{(j,k)\in\mathcal{A}(\epsilon)\cap\HH_0} [G(t)]^{-\frac{2|\rho_{jk}|}{1+|\rho_{jk}|}}.
\]
Note that for $0\le t\le b_p$, we have $G(t) \ge G(b_p) = c_p/p$, so that by assumption (A3) it follows that for some $\epsilon,q>0$,
\[
I_{11}(t) \le C\sum_{(j,k)\in\mathcal{A}(\epsilon)\cap\HH_0} p^{\frac{2|\rho_{jk}|}{1+|\rho_{jk}|}+q-2} = O(1/(\log p)^2).
\]
By the above inequalities, we can prove (\ref{5.15}) by choosing $0<\delta<1$ so that
\begin{align*}
\sum_{i=0}^{d_p}\E[I(t_i)]^2 &\le C\sum_{i=0}^{d_p}(pG(t_i))^{-1}+Cd_p[(\log p)^{-1-\delta}+(\log p)^{-2}]\\
&\le C\sum_{i=0}^{d_p} \frac{1}{c_p+c_p^{2/3}e^{i^{\delta}}}+o(1)\\
&=o(1).
\end{align*}
Lately, as all the above arguments are conditional on $\{\xx_i\}_{i=1}^n$, the statements of Theorem~\ref{fdr} follow by averaging over the probability measure of $\{\xx_i\}_{i=1}^n$ .
\qed


\section*{FUNDING}
This research was supported by NIH grants R01GM123056 and R01GM129781 and NSF grant DMS-1712735.

\section*{SUPPLEMENTARY MATERIALS}

In the Supplemental Materials, we prove all the main theorems and  the technical lemmas. 

\setstretch{1.24}
\bibliographystyle{chicago}
\bibliography{reference}

\end{document}